\documentclass[12pt]{article}
\usepackage{amsmath,amssymb,bm,epsfig,color,graphicx,cite}
\renewcommand{\thefootnote}{\fnsymbol{footnote}}

\begin{document}

\title{
\begin{flushright}
\begin{minipage}{0.2\linewidth}
\normalsize
\end{minipage}
\end{flushright}
{\Large \bf 
A precision calculation of relic neutrino decoupling
\\*[20pt]}}

\author{
Kensuke Akita$^{1}$\footnote{
E-mail address: kensuke@th.phys.titech.ac.jp}\ and\ 
Masahide Yamaguchi$^{1}$\footnote{
E-mail address: gucci@phys.titech.ac.jp}\\*[20pt]
$^1${\it \normalsize
Department of Physics, Tokyo Institute of Technology,
Tokyo 152-8551, Japan} \\*[50pt]
}

\date{
\centerline{\small \bf Abstract}
\begin{minipage}{0.9\linewidth}
\medskip \medskip \small We study the distortions of equilibrium spectra
of relic neutrinos due to the interactions with electrons,
positrons, and neutrinos in the early Universe. We solve the
integro-differential kinetic equations for the neutrino density matrix,
including three-flavor oscillations and finite temperature corrections
from QED up to the next-to-leading order $\mathcal{O}(e^3)$ for the
first time. In addition, the equivalent kinetic equations in the mass
basis of neutrinos are directly solved, and we numerically evaluate the
distortions of the neutrino spectra in the mass basis as well, which can
be easily extrapolated into those for non-relativistic neutrinos in the
current Universe. In both bases, we find the same value of the
effective number of neutrinos, $N_{\rm eff}=3.044$, which parameterizes
the total neutrino energy density. 
The estimated error for the value of $N_{\rm eff}$ due to the
 numerical calculations and the choice of neutrino mixing parameters
 would be at most $0.0005$.
\end{minipage}
}

\maketitle
\thispagestyle{empty}
\clearpage
\tableofcontents
\clearpage

\renewcommand{\thefootnote}{\arabic{footnote}}
\setcounter{footnote}{0}
%\vspace{35pt}

\section{Introduction}

The successful hot big bang model after inflation predicts that
neutrinos produced in the early Universe still exist in the current
Universe. These relic neutrinos are confirmed indirectly by the
observations of primordial abundances of light elements from Big Bang
Nucleosynthesis (BBN), the anisotropies of the Cosmic Microwave
Background (CMB) and the distribution of Large Scale Structure (LSS) of
the Universe.

The cosmic neutrino background was generated at high temperature and
kept in thermal equilibrium through weak interactions. When the
temperature of the Universe decreased, weak interactions became
ineffective and cosmic neutrinos were decoupled with other particles at
the decoupling temperature $T_{\rm dec} \sim 2 \ {\rm MeV}$. In the
instantaneous decoupling limit, the energy spectrum of neutrinos
takes a form of Fermi-Dirac distribution function and
receives only the effect of redshift of physical momentum after the
decoupling. Soon after the decoupling of neutrinos, electrons and
positrons start to annihilate and to heat photons when the temperature
of the Universe is almost equal to the electron mass $m_e = 0.511\ {\rm
MeV}$. If we assume that electrons and positrons annihilate only into
photons, we can approximately estimate the ratio of
the temperatures of cosmic photons and neutrinos, $T_{\gamma}/T_{\nu}
\simeq 1.40102$, using the entropy conservation of the Universe.

However, the decoupling temperature of neutrinos $T_{\rm dec}$ and the
temperature of annihilation of electrons and positrons are so close that
some neutrinos keep interacting with electrons and positrons. These
interaction processes become more efficient for neutrinos with higher
energies because the interaction rates of relativistic particles with
higher energies are larger\cite{Dicus:1982bz,Dolgov:2002wy}. Due to
these processes, non-thermal distortions in the neutrino spectra are
produced and the photon temperature increases less than that in the
instantaneous decoupling limit.  In particular, the
non-thermal distortions increase the total energy density of neutrinos,
which can be parameterized by the effective number of neutrinos $N_{\rm
eff}$. This parameter can be constrained by cosmological
observations such as the measurement of the CMB anisotropies.

The non-instantaneous decoupling of neutrinos was studied numerically
for a long time. In particular, this numerical study requires solving
integro-differential kinetic equations, which correspond to the
Boltzmann equations for neutrino momentum distributions.
\footnote{Recently, various approximate evaluations on $N_{\rm eff}$ are also developing. In refs. \cite{Escudero:2018mvt, Escudero:2020dfa}, a simple evaluation model of $N_{\rm eff}$ is proposed, including finite temperature QED corrections and neutrino oscillations. In ref. \cite{Bennett:2019ewm}, the authors approximately estimate $N_{\rm eff}$ with including QED corrections up to $\mathcal{O}(e^4)$.} 
First, several studies
\cite{Dodelson:1992km, Dolgov:1992qg, Fields:1992zb} solved these
Boltzmann equations under some approximations such as Maxwell-Boltzmann
statistics approximation for neutrinos. A few years later, the Boltzmann
equations for the distorted Fermi-Dirac statistics of neutrinos were
solved in refs. \cite{Hannestad:1995rs, Dolgov:1997mb,Dolgov:1998sf,
Esposito:2000hi}. Finally, the kinetic equations were solved with
including finite temperature radiative corrections at leading order
\cite{Fornengo:1997wa, Mangano:2001iu, Birrell:2014uka, Grohs:2015tfy,
Grohs:2017iit, Froustey:2019owm} and three-flavor neutrino oscillations
\cite{Mangano:2005cc, deSalas:2016ztq, Gariazzo:2019gyi}.  The
kinetic equations including neutrino oscillations correspond to the
Boltzmann equations for the neutrino density matrix. In
refs. \cite{Mangano:2005cc,Gariazzo:2019gyi}, the authors solved the
Boltzmann equations under the damping approximation, where the
off-diagonal parts of the collision terms are treated as the damping
factors. In ref. \cite{ deSalas:2016ztq}, the Boltzmann equations for
the neutrino density matrix with the full collision terms were solved.
 
In the present Universe, since the average magnitude of momenta for
neutrinos is $\langle p \rangle\sim 0.53~{\rm meV} \ll \sqrt{\Delta
m^2_{21}}, \sqrt{|\Delta m^2_{31}|}$, two massive neutrinos at least are
non-relativistic.  In the non-relativistic epoch for neutrinos, we
cannot quantize flavor neutrinos, which are flavor eigenstates of
neutrinos, in the conventional way as we quantize fields whose masses
are diagonalized, and hence the flavor neutrino spectra in this epoch do
not make sense. In order to investigate and to detect the neutrino
spectra in the present epoch, we need to consider massive neutrino
spectra. Then, in this paper, by solving the kinetic equations for
massive neutrinos in the early Universe, we study the distortions for
neutrino spectra in the mass basis too, which could be easily
extrapolated to those in the current epoch, and compare the results both
in the flavor and mass bases.  In addition, since we can expect the
better accuracy on future measurements of $N_{\rm eff}$, we solve the
Boltzmann equations for the neutrino density matrix with full collision
terms, including finite temperature corrections from QED up to the
next-to-leading order $\mathcal{O}(e^3)$ for the first time.

This paper is organized as follows. In Sec. \ref{sec:2}, we give the
Boltzmann equations both in the flavor and mass bases in order to
analyze the decoupling process of neutrinos. In this section, we also
discuss finite temperature corrections from QED and comment on the
computational method and initial conditions. In Sec. \ref{sec:3}, we
present our results of neutrino spectra and the value of $N_{\rm eff}$
for each basis. We also discuss the relation of distribution functions
in the flavor and mass bases. Finally, we give our conclusions in Sec.\
\ref{sec:con}. In appendices, the kinetic equations for neutrinos in
comoving variables and analytic estimation of the collision integral are
given.

%%%%%%%%%%%%%%%%%%%%%%%%%%%%%%%%%%%%%%%%%%%%%%%%%%%%%%%%%%%%%%%%%%%%%%%%%%%%%%
 %%%%%%%%%%%%%%%%%%%%%%%%%%%%%%%%%%%%%%%%%%%%%%%%%%%%%%%%%%%%%%%%%%%%%%%%%%%%%%
%%%%%%%%%%%%%%%%%%%%%%%%%%%%%%%%%%%%%%%%%%%%%%%%%%%%%%%%%%%%%%%%%%%%%%%%%%%%%%
\section{Neutrino decoupling} 
\label{sec:2}

\subsection{Boltzmann equations in the flavor basis}
\label{sec:2.1}

In order to describe the process of neutrino decoupling in the early
Universe, in particular, to estimate the spectral distortion with good
precision, we first consider field operators of flavor neutrinos and their
density matrices in a homogeneous system. In ultra-relativistic limit, the
field operators of left-handed neutrinos are
expanded as
%\begin{align}
%\nu_\alpha(x) = \int \frac{d^3{\bm p}}{(2\pi)^3}\left(a_\alpha({\bm{p}},t)u_{\bm{p}}+b_\alpha^{\dag}({-\bm{p}},t)v_{-\bm{p}} \right)e^{-i\bm{p\cdot x}},
%\end{align} 
\begin{align}
\bm{\nu}_\alpha(x) = \int \frac{d^3{\bm
 p}}{(2\pi)^3 \sqrt{2p_0}}\left(a_\alpha({\bm{p}},t)u_{\bm{p}}e^{i\bm{p\cdot x}} 
+b_\alpha^{\dag}({\bm{p}},t)v_{\bm{p}} e^{-i\bm{p\cdot x}} \right),
\label{operator}
\end{align} 
where $a_\alpha(\bm{p},t) = a_\alpha(\bm{p}) e^{-ip_0 t}$ and
$b_\alpha(\bm{p},t) = b_\alpha(\bm{p}) e^{-ip_0 t}$ are annihilation
operators for negative-helicity neutrinos and positive-helicity anti-neutrinos,
respectively. $\alpha$ and $\bm{p}$ are a flavor index and a three
dimensional momentum with $p_0 \simeq |\bm{p}|$,
respectively. $u_{\bm p}\ (v_{\bm p})$ denotes the Dirac spinor for a
massless negative-helicity particle (positive-helicity anti-particle), which is
normalized to be %unity,
$u_{\bm{p}}^{\dagger}u_{\bm{p}} = v_{\bm{p}}^{\dagger}v_{\bm{p}} = 2
p_0$,
and satisfies
\begin{align}
/\hspace{-2.2mm}pu_{\bm p}=0,\ \ \ \ \ \ \ \ /\hspace{-2.2mm}pv_{\bm p}=0.
\end{align}
These expansions of the field operators make sense only in the
ultra-relativistic limit. The annihilation and creation operators satisfy the anti-commutation relations,
\begin{align}
\{ a_\alpha(\bm{p}), a_\beta^{\dag}(\bm{p}') \} = \{b_\alpha(\bm{p}), b_\beta^{\dag}(\bm{p}') \} = \delta_{\alpha\beta}(2\pi)^3\delta^{(3)}(\bm{p}-\bm{p}').
\end{align}
The density matrices for neutrinos and anti-neutrinos are defined
through the following expectation values of these operators
with regard to the initial thermal equilibrium states,
\begin{align}
\langle a^{\dag}_\beta(\bm{p},t)a_\alpha(\bm{p}',t) \rangle &= (2\pi)^3\delta^{(3)}(\bm{p}-\bm{p'})\left(\rho_p\right)_{\alpha\beta}, \nonumber \\
\langle b^{\dag}_\alpha(\bm{p},t)b_\beta(\bm{p'},t) \rangle &= (2\pi)^3\delta^{(3)}(\bm{p}-\bm{p'})\left(\bar{\rho}_p\right)_{\alpha\beta},
\label{DM}
\end{align}
where $p=|\bm{p}|$. Due to the reversed order of flavor indices in
$\bar{\rho}_p(t)$, both density matrices transform in the same way under
a unitary transformation of flavor space. Here the diagonal parts are
the usual distribution functions of flavor neutrinos and the
off-diagonal parts are non-zero in the presence of flavor mixing. Note
again that since the off-diagonal parts of the neutrino mass matrix
are not zero, we cannot define annihilation operators, creation
operators and density matrices for non-relativistic flavor neutrinos in
the conventional way. In Sec. \ref{sec:2.2}, we write field operators of
neutrinos including their masses in the mass basis so that we can define
density matrices for non-relativistic neutrinos.

The neutrino density matrix takes the following form,
\begin{align}
\rho_p(t)=
\begin{pmatrix}
\rho_{ee} & \rho_{e\mu} & \rho_{e\tau} \\
\rho_{\mu e} & \rho_{\mu \mu} & \rho_{\mu \tau} \\
\rho_{\tau e} & \rho_{\tau \mu} & \rho_{\tau \tau} 
\end{pmatrix}
=
\begin{pmatrix}
f_{\nu_e} & a_1+ia_2 & b_1+ib_2 \\
a_1-ia_2 & f_{\nu_\mu} & c_1+ic_2 \\
b_1-ib_2 & c_1-ic_2 & f_{\nu_\tau} 
\end{pmatrix},
\end{align}
where $f_{\nu_\alpha}$ is the distribution function for flavor neutrinos and the off-diagonal parts are characterized by the real parameters $a_i, b_i$ and $c_i\ (i = 1,2)$.
Hereafter we neglect a neutrino asymmetry since neutrino
oscillations leading to flavor equilibrium before BBN impose a stringent
constraint on this asymmetry \cite{Dolgov:2002ab, Wong:2002fa, Abazajian:2002qx, Mangano:2011ip,
Castorina:2012md}. Under this assumption, neutrinos and anti-neutrinos
satisfy the same density matrices and the same evolutions in the
Universe,\ $\rho_p(t) = \bar{\rho}_p(t)^{\mathrm{T}}$.

The equations of motion for the neutrino density matrix in the expanding Universe are \cite{Sigl:1992fn, Blaschke:2016xxt}
\begin{align}
(\partial_t-Hp\partial_p)\rho_p(t) = -i\left[  \left(\frac{M^2}{2p} -\frac{8\sqrt{2}G_Fp}{3m_W^2}E \right),\ \rho_p(t)  \right] +C[\rho_p(t)],
\label{BE}
\end{align}
where $H$ is the Hubble parameter, $G_F$ is the Fermi coupling constant,
$m_W$ is the W boson mass, and $[\cdot, \cdot]$ represents the
commutator of matrices with a flavor index. The first
term\footnote{Only when we derive the first term which comes from the
free neutrino Hamiltonian including the mass matrix, we replace the
operators $a_{\alpha}(\bm{p},t)$ and $b_{\alpha}(\bm{p},t)$ in
Eq.~(\ref{operator}) with $a_{\alpha}^{\rm
osc}(\bm{p},t)=(\exp(-i\Omega_{\bm{p}}t))_{\alpha\beta}a_{\beta}(\bm{p})$
and $b_{\alpha}^{\rm
osc}(\bm{p},t)=(\exp(-i\Omega_{\bm{p}}t))_{\alpha\beta}b_{\beta}(\bm{p})$
as in \cite{Sigl:1992fn}, where $\Omega_{\bm{p}}=\sqrt{\bm{p}^2+M^2}$. }
in the commutator is the vacuum oscillation term proportional to the
mass-squared matrix in the flavor basis $M^2$. The mass-squared matrix
is related to the diagonal mass-squared matrix in the mass basis $M_{\rm
diag}^2$ through the Pontecorvo-Maki-Nakagawa-Sakata matrix, assuming
the CP conservation,
\begin{align}\label{eq:PMNS}
U_{\rm PMNS} &\equiv
\begin{pmatrix}
U_{e1} & U_{e2} & U_{e3} \\
U_{\mu 1} & U_{\mu 2} & U_{\mu 3} \\
U_{\tau 1} & U_{\tau 2} & U_{\tau 3} 
\end{pmatrix}, \nonumber \\
&=
\begin{pmatrix}
c_{12}c_{13} & s_{12}c_{13} & s_{13} \\
-s_{12}c_{23}-c_{12}s_{23}s_{13} & c_{12}c_{23}-s_{12}s_{23}s_{13} & s_{23}c_{13} \\
s_{12}s_{23}-c_{12}c_{23}s_{13} & -c_{12}s_{23}-s_{12}c_{23}s_{13} & c_{23}c_{13} 
\end{pmatrix},
\end{align}
where $c_{ij} =\cos\theta_{ij}$ and $s_{ij}=\sin\theta_{ij}$ for $ij = 12,\ 13$, or $23$. The relation between the mass-squared matrices of neutrinos in the two bases is 
\begin{align}
M_{\rm diag}^2 &= U_{\rm PMNS}^{\dag}M^2U_{\rm PMNS}, \nonumber \\
&= {\rm diag}(m_1^2,\ m_2^2,\ m_3^2).
\end{align}
From the global analysis of neutrino oscillation experiments in \cite{Esteban:2018azc}, we consider the following best-fit values of neutrino masses and mixing parameters,
\begin{align}
\left(\frac{\Delta m^2_{21}}{10^{-5}\ {\rm eV^2}},\ \frac{\Delta m_{31}^2}{10^{-3}\ {\rm eV^2}},\ s_{12}^2,\ s_{23}^2,\ s_{13}^2 \right)_{\rm NH}
&= (7.39,\ 2.525,\ 0.310,\ 0.582,\ 0.0224), \nonumber \\
\left(\frac{\Delta m^2_{21}}{10^{-5}\ {\rm eV^2}},\ \frac{\Delta m_{31}^2}{10^{-3}\ {\rm eV^2}},\ s_{12}^2,\ s_{23}^2,\ s_{13}^2 \right)_{\rm IH}
&= (7.39,\ -2.512,\ 0.310,\ 0.582,\ 0.02263),
\label{VN}
\end{align}
where $\Delta m^2_{ij}=m^2_i-m^2_j$. The first (second) equation in
Eq.~(\ref{VN}) corresponds to the normal (inverted) hierarchy ordering of
neutrino masses.

The second term in the commutator in Eq.~(\ref{BE}) represents the
refractive effect in the medium which comes from one-loop thermal
contributions to the neutrino self-energy. The diagonal matrix $E$ is
the energy density of the charged leptons and in the temperature of
${\rm MeV}$ scale, $E$ takes the following form approximately,
\begin{align}
E = {\rm diag}(\rho_{ee},\ 0,\ 0),
\end{align}
where $\rho_{ee}=\rho_{e^-}+\rho_{e^+}$ is the energy density of
electrons and positrons. We neglect other refractive terms coming from
the charged lepton asymmetries and neutrino self-interactions, which are
significantly suppressed \cite{Sigl:1992fn, Notzold:1987ik}.

The final term in Eq.~(\ref{BE}) represents the collisions of neutrinos
with electrons, positrons, and themselves, which are dominated by
two-body reactions $1+2\rightarrow3+4$.
As done in the previously most accurate calculation of neutrino decoupling in the
early Universe \cite{deSalas:2016ztq}, we also deal with both diagonal and
off-diagonal collision terms for the processes which involve electrons
and positrons. On the other hand, we do not treat the off-diagonal terms
for the self-interactions of neutrinos, such as $\nu\nu \leftrightarrow
\nu\nu$ or $\nu\bar{\nu} \leftrightarrow \nu\bar{\nu}$ since
the annihilations of electrons and positrons are important for the
heating process of neutrinos while the self-interactions of neutrinos less contribute
to this heating process.

The diagonal collision term from the self-interaction processes
$\nu(p_1)\nu(p_2) \leftrightarrow \nu(p_3)\nu(p_4)$ and
$\nu(p_1)\bar{\nu}(p_2) \leftrightarrow
\nu(p_3)\bar{\nu}(p_4)$ is
\begin{align}
C_S[\nu_\alpha(p_1)] &= \frac{2^5G_F^2}{2\left|\bm{p}_1\right|}\int \frac{d^3\bm{p}_2}{(2\pi)^32\left|\bm{p}_2\right|}\frac{d^3\bm{p}_3}{(2\pi)^32\left|\bm{p}_3\right|}\frac{d^3\bm{p}_4}{(2\pi)^32\left|\bm{p}_4\right|}(2\pi)^4\delta^{(4)}(p_1+p_2-p_3-p_4) \nonumber \\
&\times \Bigl[ \left\{4(p_1\cdot p_4)(p_2\cdot p_3)+2(p_1\cdot p_2)(p_3\cdot p_4) \right\}F(\nu_\alpha ^{(1)}, \nu_\alpha^{(2)}, \nu_\alpha^{(3)}, \nu_\alpha^{(4)}  )  \nonumber \\
&+  \left\{(p_1\cdot p_4)(p_2\cdot p_3)+(p_1\cdot p_2)(p_3\cdot p_4) \right\}F(\nu_\alpha^{(1)}, \nu_\beta^{(2)}, \nu_\alpha^{(3)}, \nu_\beta^{(4)} ) \nonumber \\
&+ (p_1\cdot p_4)(p_2\cdot p_3)F\left(\nu_\alpha^{(1)}, \nu_\alpha^{(2)}, \nu_\beta^{(3)}, \nu_\beta^{(4)}  \right) \nonumber \\
&+  \left\{(p_1\cdot p_4)(p_2\cdot p_3)+(p_1\cdot p_2)(p_3\cdot p_4) \right\}F(\nu_\alpha^{(1)}, \nu_\gamma^{(2)}, \nu_\alpha^{(3)}, \nu_\gamma^{(4)} ) \nonumber \\
&+ (p_1\cdot p_4)(p_2\cdot p_3)F(\nu_\alpha^{(1)}, \nu_\alpha^{(2)}, \nu_\gamma^{(3)}, \nu_\gamma^{(4)} )
\Bigl],
\end{align}
where $\alpha,\beta,\gamma = e,\mu,\tau$ and $\alpha\neq \beta, \alpha\neq \gamma, \beta\neq \gamma$. We define $F(\nu_\alpha ^{(1)}, \nu_\beta^{(2)}, \nu_\gamma^{(3)}, \nu_\delta^{(4)} )$ as
\begin{align}
F(\nu_\alpha ^{(1)}, \nu_\beta^{(2)}, \nu_\gamma^{(3)}, \nu_\delta^{(4)} ) &= f_{\nu_\gamma}(p_3)f_{\nu_\delta}(p_4)\left(1-f_{\nu_\alpha}(p_1)\right)\left(1-f_{\nu_\beta}(p_2)\right) \nonumber \\
&-f_{\nu_\alpha}(p_1)f_{\nu_\beta}(p_2)\left(1-f_{\nu_\gamma}(p_3)\right)\left(1-f_{\nu_\delta}(p_4)\right),
\label{FS}
\end{align}
where $\alpha,\beta,\delta,\gamma = e,\mu,\tau$.

The collision term from the annihilation processes $\nu(p_1)\bar{\nu}(p_2) \leftrightarrow e^-(p_3)e^+(p_4)$ is
\begin{align}
C_A &= \frac{1}{2}\frac{2^5G_F^2}{2\left|\bm{p}_1\right|}\int \frac{d^3\bm{p}_2}{(2\pi)^32\left|\bm{p}_2\right|}\frac{d^3\bm{p}_3}{(2\pi)^32E_3}\frac{d^3\bm{p}_4}{(2\pi)^32E_4}(2\pi)^4\delta^{(4)}(p_1+p_2-p_3-p_4) \nonumber \\
&\times \Bigl[ 4(p_1\cdot p_4)(p_2\cdot p_3)F^{LL}_A\left(\nu^{(1)}, \bar{\nu}^{(2)}, e^{(3)}, \bar{e}^{(4)}\right) \nonumber \\
&+ 4(p_1\cdot p_3)(p_2\cdot p_4)F^{RR}_A\left(\nu^{(1)}, \bar{\nu}^{(2)}, e^{(3)}, \bar{e}^{(4)}\right) \nonumber \\
&+2(p_1\cdot p_2)m_e^2 \Bigl(F^{LR}_A\left(\nu^{(1)}, \bar{\nu}^{(2)}, e^{(3)}, \bar{e}^{(4)}\right) + F^{RL}_A\left(\nu^{(1)}, \bar{\nu}^{(2)}, e^{(3)}, \bar{e}^{(4)}\right) \Bigl)
\Bigl],
\end{align}
where
\begin{align}
&F^{ab}_A\left(\nu^{(1)}, \bar{\nu}^{(2)}, e^{(3)}, \bar{e}^{(4)}\right) \nonumber \\
&= f_{e}(p_3)f_{e}(p_4)
\Bigl( Y^a(1-\bar{\rho}_2)Y^b(1-\rho_1)+(1-\rho_1)Y^b(1-\bar{\rho}_2)Y^a \Bigl) \nonumber \\
&-(1-f_{e}(p_3))(1-f_{e}(p_4))\Bigl( Y^a\bar{\rho}_2Y^b\rho_1+\rho_1Y^b\bar{\rho}_2Y^a \Bigl).
\label{FA}
\end{align}
Here $f_{e}(p)$ is the distribution functions of electrons and positrons,
and $\bar{\rho}_2= \rho_2^{\mathrm{T}}$. We assume that electrons and positrons are always in thermal
equilibrium since electrons, positrons and photons interact with each
other through rapid electromagnetic interactions. Under this assumption, the
distribution functions of electrons and positrons take the following
form,
\begin{align}
f_e(p) = \frac{1}{\exp(\sqrt{p^2+m_e^2}/T_{\gamma})+1}.
\end{align}
$Y^a (a = L, R)$ is a $3 \times 3$ matrix of couplings and becomes in
the flavor basis
\begin{align}
Y^L &= {\rm diag}(g_L,\tilde{g}_L,\tilde{g}_L), \nonumber \\
Y^R &= {\rm diag}(g_R,g_R,g_R),
\end{align}
where
\begin{align}
g_L = \frac{1}{2}+\sin^2\theta_W,\ \ \ \   \tilde{g}_L = -\frac{1}{2}+\sin^2\theta_W,\ \ \ \ g_R = \sin^2\theta_W.
\end{align}
Here $\sin^2\theta_W \simeq 0.231$ and $\theta_W$ is the weak mixing
angle.

The collision term from the scattering processes $\nu(p_1)e^-(p_2) \leftrightarrow \nu(p_3)e^-(p_4)$ and \\
$\nu(p_1)e^+(p_2) \leftrightarrow \nu(p_3)e^+(p_4)$ is 
\begin{align}
C_{SC} &= \frac{1}{2}\frac{2^5G_F^2}{2\left|\bm{p}_1\right|}\int \frac{d^3\bm{p}_2}{(2\pi)^32E_2}\frac{d^3\bm{p}_3}{(2\pi)^32\left|\bm{p}_3\right|}\frac{d^3\bm{p}_4}{(2\pi)^32E_4}(2\pi)^4\delta^{(4)}(p_1+p_2-p_3-p_4) \nonumber \\
&\times  \Bigl[  4\left\{(p_1\cdot p_4)(p_2\cdot p_3)+(p_1\cdot p_2)(p_3\cdot p_4) \right\} \nonumber \\
&\times \Bigl(F^{LL}_{SC}\left(\nu^{(1)},e^{(2)}, \nu^{(3)},e^{(4)}\right) + F^{RR}_{SC}\left(\nu^{(1)},e^{(2)}, \nu^{(3)},e^{(4)}\right) \Bigl) \nonumber \\
&-4(p_1\cdot p_3)m_e^2  \Bigl(F^{LR}_{SC}\left(\nu^{(1)},e^{(2)}, \nu^{(3)},e^{(4)}\right) + F^{RL}_{SC}\left(\nu^{(1)},e^{(2)}, \nu^{(3)},e^{(4)}\right) \Bigl)
\Bigl],
\end{align}
where
\begin{align}
&F^{ab}_{SC}\left(\nu^{(1)},e^{(2)}, \nu^{(3)},e^{(4)}\right) \nonumber \\
&= f_e(p_4)(1-f_e(p_2))\Bigl(Y^a\rho_3Y^b(1-\rho_1) + (1-\rho_1)Y^b\rho_3Y^a \Bigl) \nonumber \\
&-f_e(p_2)(1-f_e(p_4))\Bigl( \rho_1Y^b(1-\rho_3)Y^a+Y^a(1-\rho_3)Y^b\rho_1 \Bigl).
\label{FSC}
\end{align}
These collision terms are described in detail in appendix \ref{appa}.

In addition to the Boltzmann equations for the neutrino density
matrix, the energy conservation law must be satisfied,
\begin{align}
\frac{d\rho}{dt}=-3H(\rho+P),
\label{EC}
\end{align}
where $\rho$ and $P$ are the total energy density and pressure of the
standard model particles ($\gamma, e^{\pm}, \nu_i$)
respectively. Though we will discuss finite temperature corrections
from QED to $\rho, P$ and $m_e$ later, in
the ideal gas limit, they are given as follows, which are denoted by
$\rho_{(0)}$ and $P_{(0)}$ respectively,
\begin{align}
\rho_{(0)} &= \frac{\pi^2T_{\gamma}^4}{15}+\frac{2}{\pi^2}\int\frac{dpp^2\sqrt{p^2+m_e^2}}{\exp(\sqrt{p^2+m_e^2}/T_{\gamma})+1}+\sum_{\alpha=e,\mu,\tau}\frac{1}{\pi^2}\int dp~p^3f_{\nu_\alpha}(p), \nonumber \\
P_{(0)} &= \frac{\pi^2T_{\gamma}^4}{45}+\frac{2}{\pi^2}\int\frac{dpp^4}{3\sqrt{p^2+m_e^2}[\exp(\sqrt{p^2+m_e^2}/T_{\gamma})+1]}+\sum_{\alpha=e,\mu,\tau}\frac{1}{3\pi^2}\int dp~p^3f_{\nu_\alpha}(p).
\end{align}
The energy conservation law governs the evolution of the photon
temperature $T_{\gamma}$. The Hubble parameter in Eqs.~(\ref{BE}) and
(\ref{EC}) is calculated using the usual relation,
$3H^2m_{\rm Pl}^2=8\pi\rho$ with $m_{\rm Pl}$ being the Planck mass, where
we ignore the curvature term and the cosmological constant because they
are negligible in the radiation dominated epoch.

%%%%%%%%%%%%%%%%%%%%%%%%%%%%%%%%%%%%%%%%%%%%%%%%%%%%%%%%%%%%%%%%%%%%%%%%%%%%%%%%%%%%%%%%%%%%%%%%%%
\subsection{Boltzmann equations in the mass basis}
\label{sec:2.2}

In this section, we formulate the Boltzmann equations for the
density matrices of massive neutrinos at the early Universe in the
mass basis.
If we would like to observe
the distortions of neutrinos in the current Universe in future, it is
easier to follow the evolution of negative-helicity neutrinos in the diagonal mass basis
since the helicity states of neutrinos are conserved while
non-relativistic neutrinos are freely streaming. Thus, it is quite
useful to formulate the Boltzmann equations for negative-helicity neutrinos in the mass basis though
we concentrate on neutrino decoupling processes in this
paper, where ultra-relativistic limit is a good approximation and there
are no much difference between the two bases.
This approach is also complementary to that in the flavor
basis given in the previous subsection \ref{sec:2.1} and is useful for
the cross-check of the results.

Since the negative-helicity neutrinos in the mass basis satisfy the free
Dirac equation, they are expanded as
%\begin{align}
%\tilde{\nu}_i(x) = \int \frac{d^3{\bm p}}{(2\pi)^3}\left(a_i({\bm{p}},t)\tilde{u}_{\bm{p}}e^{iE_it-i\bm{p\cdot x}}+b_i^{\dag}({\bm{p}},t)\tilde{v}_{\bm{p}}e^{-iE_it+i\bm{p\cdot x}} \right),
%\end{align}
\begin{align}
\bm{\nu}_i(x) = \int \frac{d^3{\bm p}}{(2\pi)^3 \sqrt{2E_i}}\left(a_i({\bm{p}},t)u^{(i)}_{\bm{p}}e^{i\bm{p\cdot x}}+b_i^{\dag}({\bm{p}},t)v^{(i)}_{\bm{p}}e^{-i\bm{p\cdot x}} \right),
\end{align}
where $i(=1,2,3)$ represents a mass eigenstate, $a_i({\bm{p}},t) =
a_i({\bm{p}}) e^{-iE_it}, b_i({\bm{p}},t) = b_i({\bm{p}}) e^{-iE_it}$,
$E_i = \sqrt{\bm{p}^2+m_i^2}$ and $m_i$ is the neutrino mass in the
mass basis. Since $u^{(i)}_{\bm p}\ (v^{(i)}_{\bm p})$
denotes the Dirac spinor for massive negative-helicity particles\
(positive-helicity anti-particles), which is also normalized to be
$u^{(i)}_{\bm{p}}{}^{\dagger}u^{(i)}_{\bm{p}} = 
v^{(i)}_{\bm{p}}{}^{\dagger}v^{(i)}_{\bm{p}} = 2E_i$,
the Dirac spinors satisfy
\begin{align}
(/\hspace{-2.2mm}p-m_i)u^{(i)}_{\bm p}=0,\ \ \ \ \ \ \ \ (/\hspace{-2.2mm}p+m_i)v^{(i)}_{\bm p}=0.
\end{align}
As in the flavor basis, $a_i(\bm{p})$ and $b_i(\bm{p})$ are
annihilation operators for negative-helicity neutrinos and for positive-helicity
anti-neutrinos in the mass basis, respectively, which satisfy
\begin{align}
\{ a_i(\bm{p}), a_j^{\dag}(\bm{p}') \} = \{b_i(\bm{p}), b_j^{\dag}(\bm{p}') \} = \delta_{ij}(2\pi)^3\delta^{(3)}(\bm{p}-\bm{p}').
\end{align}
The density matrices for neutrinos and anti-neutrinos in the mass basis are given by
\begin{align}
\langle a^{\dag}_j(\bm{p},t)a_i(\bm{p}',t) \rangle &= (2\pi)^3\delta^{(3)}(\bm{p}-\bm{p'})\left(\rho_p\right)_{ij}, \nonumber \\
\langle b^{\dag}_i(\bm{p},t)b_j(\bm{p'},t) \rangle &= (2\pi)^3\delta^{(3)}(\bm{p}-\bm{p'})\left(\bar{\rho}_p\right)_{ij},
\label{DM2}
\end{align}
where the diagonal parts are the distribution functions for massive
neutrinos.

In the following, we study the kinetic equations for the neutrino density matrix in ultra-relativistic limit.
In this limit, negative-helicity neutrinos coincide with left-handed neutrinos due to no distinction between helicity and chirality.
The diagonalization of mass matrix for left-handed neutrinos in the flavor basis is achieved through the transformations,
\begin{align}
\bm{\nu}_{\alpha}(x) = \sum_{i=1}^3U_{\alpha i}\bm{\nu}_i(x),
\end{align}
with $\alpha =e,\mu,\tau$. Here $U_{\alpha i}$ represents a component of
the unitary matrix $U_{\rm PMNS}$ given in Eq.~(\ref{eq:PMNS}).

In order to specify the collision processes for massive neutrinos, we
discuss weak neutral currents and charged currents in the mass basis. The
weak neutral currents of electrons, positrons, and neutrinos are
\begin{align}
J_{\rm NC}^{\mu}=J_{ee}^{L\mu}+J_{ee}^{R\mu}+J_{\nu\nu}^{\mu},
\end{align}
where $J_{ee}^{L\mu}$ and $J_{ee}^{R\mu}$ are the neutral currents of left-handed electrons and right-handed electrons respectively and given by
\begin{align}
J_{ee}^{L\mu} &= \tilde{g}_L\bar{\bm{e}}\gamma^{\mu}(1-\gamma_5)\bm{e}, \nonumber \\
J_{ee}^{R\mu} &= g_R\bar{\bm{e}}\gamma^{\mu}(1+\gamma_5)\bm{e},
\end{align}
where $\bm{e}$ is the field operator of electron.  The neutral currents
for the left-handed neutrinos in both bases are given by the following
form and they are related through the unitary $U_{\rm PMNS}$ matrix,
\begin{align}
J_{\nu\nu}^\mu = \sum_{\alpha=e,\mu,\tau} \bar{\bm{\nu}}_\alpha\gamma^\mu(1-\gamma_5)\bm{\nu}_\alpha
= \sum_{i = 1,2,3} \bar{\bm{\nu}}_i\gamma^\mu(1-\gamma_5)\bm{\nu}_i.
\end{align} 

The charged currents for left-handed electrons and left-handed electron neutrinos are given by
\begin{align}
J_{e\nu_e}^\mu=\bar{\bm{\nu}}_e\gamma^\mu(1-\gamma_5)\bm{e}.
\end{align}
Using the Fierz transformations for fermionic fields, we can describe the Hamiltonian density including the charged currents at the neutrino decoupling process as
\begin{align}
\mathcal{H}_{CC} = \frac{G_F}{\sqrt{2}}  J^{\dag\mu}_{e\nu_e}(J_{e\nu_e})_\mu &= \frac{G_F}{\sqrt{2}}J_{ee}^{L\mu}(J_{\nu_e\nu_e})_\mu, \nonumber \\
J_{\nu_e\nu_e}^\mu &= \bar{\bm{\nu}}_e\gamma^\mu(1-\gamma_5)\bm{\nu}_e.
\label{FI}
\end{align}
Since Eq.~(\ref{FI}) implies that the charged currents in the Hamiltonian density $\mathcal{H}_{CC}$
can be replaced by the equivalent neutral currents, we implicitly take the
charged currents into account by making the following replacement of the
coefficient of neutral currents for electron neutrinos in the collision
terms,
\begin{align}
\tilde{g}_L \rightarrow g_L = \tilde{g}_L+1.
\end{align}
In the mass basis, the charged currents for neutrinos and electrons are, through the $U_{\rm PMNS}$ matrix,
\begin{align}
J_{e\nu_e}^\mu&=\bar{\bm{\nu}}_e\gamma^\mu(1-\gamma_5)\bm{e}, \nonumber \\
&=\sum_{i=1}^3U^*_{ei}\bar{\bm{\nu}}_i\gamma^\mu(1-\gamma_5) \bm{e}.
\end{align}
The replacement in Eq.~(\ref{FI}) corresponds to the following relation in the mass basis,
\begin{align}
J^{\dag\mu}_{e\nu_e}(J_{e\nu_e})_\mu &= \sum_{i=1}^3\sum_{j=1}^3U_{ei}^*U_{ej} J_{ee}^{L\mu}(J_{{\nu}_i{\nu}_j})_\mu, \nonumber \\
J_{\nu_i\nu_j}^\mu &= \bar{\bm{\nu}}_i\gamma^\mu(1-\gamma_5)\bm{\nu}_j.
\end{align}
From the above equation, the corresponding couplings to $Y^a$ in
Eqs.~(\ref{FA}) and (\ref{FSC}) are changed into
\begin{align}
Y^L &\rightarrow Z^L =
\begin{pmatrix}
\tilde{g}_L + U_{e1}^*U_{e1} & U_{e1}^*U_{e2} & U_{e1}^*U_{e3} \\
U_{e2}^*U_{e1} & \tilde{g}_L + U_{e 2}^*U_{e2} & U_{e 2}^*U_{e3} \\
U_{e3}^*U_{e1} & U_{e 3}^*U_{e2} & \tilde{g}_L + U_{e 3}^*U_{e3}
\end{pmatrix}, \nonumber \\
Y^R &\rightarrow Z^R = Y^R = {\rm diag}(g_R,g_R,g_R).
\label{CM}
\end{align}
The collision term from the self-interaction processes in the mass basis
takes the same form as that in the flavor basis except for the subscripts,
\begin{align}
C_S[\nu_i(p_1)] &= \frac{2^5G_F^2}{2\left|\bm{p}_1\right|}\int \frac{d^3\bm{p}_2}{(2\pi)^32\left|\bm{p}_2\right|}\frac{d^3\bm{p}_3}{(2\pi)^32\left|\bm{p}_3\right|}\frac{d^3\bm{p}_4}{(2\pi)^32\left|\bm{p}_4\right|}(2\pi)^4\delta^{(4)}(p_1+p_2-p_3-p_4) \nonumber \\
&\times \Bigl[ \left\{4(p_1\cdot p_4)(p_2\cdot p_3)+2(p_1\cdot p_2)(p_3\cdot p_4) \right\}F(\nu_i ^{(1)}, \nu_i^{(2)}, \nu_i^{(3)}, \nu_i^{(4)}  )  \nonumber \\
&+  \left\{(p_1\cdot p_4)(p_2\cdot p_3)+(p_1\cdot p_2)(p_3\cdot p_4) \right\}F(\nu_i ^{(1)}, \nu_j^{(2)}, \nu_i^{(3)}, \nu_j^{(4)} ) \nonumber \\
&+ (p_1\cdot p_4)(p_2\cdot p_3)F\left(\nu_i ^{(1)}, \nu_i^{(2)}, \nu_j^{(3)}, \nu_j^{(4)}  \right) \nonumber \\
&+  \left\{(p_1\cdot p_4)(p_2\cdot p_3)+(p_1\cdot p_2)(p_3\cdot p_4) \right\}F(\nu_i ^{(1)}, \nu_k^{(2)}, \nu_i^{(3)}, \nu_k^{(4)} ) \nonumber \\
&+ (p_1\cdot p_4)(p_2\cdot p_3)F(\nu_i ^{(1)}, \nu_i^{(2)}, \nu_k^{(3)}, \nu_k^{(4)} )
\Bigl],
\end{align}
where $i,j,k =1,2,3$ and $i\neq j, i\neq k, j\neq k$.
The collision term from the annihilation processes is 
\begin{align}
C_A &= \frac{1}{2}\frac{2^5G_F^2}{2\left|\bm{p}_1\right|}\int \frac{d^3\bm{p}_2}{(2\pi)^32\left|\bm{p}_2\right|}\frac{d^3\bm{p}_3}{(2\pi)^32E_3}\frac{d^3\bm{p}_4}{(2\pi)^32E_4}(2\pi)^4\delta^{(4)}(p_1+p_2-p_3-p_4) \nonumber \\
&\times \Bigl[ 4(p_1\cdot p_4)(p_2\cdot p_3)G^{LL}_A\left(\nu^{(1)},\bar{\nu}^{(2)}, e^{(3)}, \bar{e}^{(4)}\right) \nonumber \\
&+ 4(p_1\cdot p_3)(p_2\cdot p_4)G^{RR}_A\left(\nu^{(1)}, \bar{\nu}^{(2)}, e^{(3)}, \bar{e}^{(4)}\right) \nonumber \\
&+2(p_1\cdot p_2)m_e^2 \Bigl(G^{LR}_A\left(\nu^{(1)}, \bar{\nu}^{(2)}, e^{(3)}, \bar{e}^{(4)}\right) + G^{RL}_A\left(\nu^{(1)}, \bar{\nu}^{(2)}, e^{(3)}, \bar{e}^{(4)}\right) \Bigl)
\Bigl],
\end{align}
where
\begin{align}
&G^{ab}_A\left(\nu^{(1)}, \bar{\nu}^{(2)}, e^{(3)}, \bar{e}^{(4)}\right) \nonumber \\
&= f_{e}(p_3)f_{e}(p_4)
\Bigl( Z^a(1-\bar{\rho}_2)Z^b(1-\rho_1)+(1-\rho_1)Z^b(1-\bar{\rho}_2)Z^a \Bigl) \nonumber \\
&-(1-f_{e}(p_3))(1-f_{e}(p_4))\Bigl( Z^a\bar{\rho}_2Z^b\rho_1+\rho_1Z^b\bar{\rho}_2Z^a \Bigl).
\label{GA}
\end{align}
 The collision term from the scattering processes is 
 \begin{align}
C_{SC} &= \frac{1}{2}\frac{2^5G_F^2}{2\left|\bm{p}_1\right|}\int \frac{d^3\bm{p}_2}{(2\pi)^32E_2}\frac{d^3\bm{p}_3}{(2\pi)^32\left|\bm{p}_3\right|}\frac{d^3\bm{p}_4}{(2\pi)^32E_4}(2\pi)^4\delta^{(4)}(p_1+p_2-p_3-p_4) \nonumber \\
&\times  \Bigl[  4\left\{(p_1\cdot p_4)(p_2\cdot p_3)+(p_1\cdot p_2)(p_3\cdot p_4) \right\} \nonumber \\
&\times \Bigl(G^{LL}_{SC}\left(\nu^{(1)},e^{(2)}, \nu^{(3)},e^{(4)}\right) + G^{RR}_{SC}\left(\nu^{(1)},e^{(2)}, \nu^{(3)},e^{(4)}\right) \Bigl) \nonumber \\
&-4(p_1\cdot p_3)m_e^2  \Bigl(G^{LR}_{SC}\left(\nu^{(1)},e^{(2)}, \nu^{(3)},e^{(4)}\right) + G^{RL}_{SC}\left(\nu^{(1)},e^{(2)}, \nu^{(3)},e^{(4)}\right) \Bigl)
\Bigl],
\end{align}
where
\begin{align}
&G^{ab}_{SC}\left(\nu^{(1)},e^{(2)}, \nu^{(3)},e^{(4)}\right) \nonumber \\
&= f_e(p_4)(1-f_e(p_2))\Bigl(Z^a\rho_3Z^b(1-\rho_1) + (1-\rho_1)Z^b\rho_3Z^a \Bigl) \nonumber \\
&-f_e(p_2)(1-f_e(p_4))\Bigl( \rho_1Z^b(1-\rho_3)Z^a+Z^a(1-\rho_3)Z^b\rho_1 \Bigl).
\label{GSC}
\end{align}
Finally, the equations of motion for the neutrino density matrix
in the mass basis are given by
\begin{align}
(\partial_t-Hp\partial_p)\rho_p(t) = -i\left[  \left(\frac{M_{\rm diag}^2}{2p} -\frac{8\sqrt{2}G_Fp}{3m_W^2}\tilde{E} \right),\ \rho_p(t)  \right] +C[\rho_p(t)].
\label{BE2}
\end{align}
Since $\tilde{E}$ and $E$ are the thermal contributions to the
self-energies of left-handed neutrinos, these have the same relation as that of $M^2_{\rm diag}$ and
$M^2$, which is given by,
\begin{align}
\tilde{E} = U_{\rm PMNS}^{\dag}EU_{\rm PMNS}.
\end{align}
 
 %%%%%%%%%%%%%%%%%%%%%%%%%%%%%%%%%%%%%%%%%%%%%%%%%%%%%%%%%%%%%%%%%%%%%%%%%%%%%%%%%%%%%%%%%%%%%%%%%%
 \subsection{Finite temperature corrections from QED}
 
In this section, we discuss finite temperature corrections from QED up
to the next-to-leading order $\mathcal{O}(e^3)$, which modify electron, positron and photon
masses. Through these corrections, several points in the former sections
are changed. First, the modification of masses affects the energy density and the pressure of the electromagnetic plasma, and the collision rates involving
electrons and positrons. In addition, the expansion rate $H$ in the
Boltzmann equations changes through the total energy density of the
plasma.
 
The corrections to electron, positron and photon masses can be obtained
perturbatively by calculating the loop corrections to the self-energies of
these particles. The corrections to the electron and positron masses
from finite temperature effects at $\mathcal{O}(e^2)$ are given by
\cite{Heckler:1994tv},
 \begin{align}
 \delta m_{e(2)}^2(p,T_{\gamma}) &= \frac{2\pi \alpha T_{\gamma}^2}{3}+\frac{4\alpha}{\pi}\int_0^{\infty}dk \frac{k^2}{E_k}\frac{1}{\exp(E_k/T_{\gamma})+1} \nonumber \\
 &-\frac{2m_e^2\alpha}{\pi p}\int^{\infty}_0dk\frac{k}{E_k}\log \left| \frac{p+k}{p-k} \right|\frac{1}{\exp(E_k/T_{\gamma})+1},
 \label{delta_me}
 \end{align}
where $\alpha = e^2/4\pi$ and $E_k = \sqrt{k^2+m_e^2}$. 
The last
term gives less than a $10\%$ correction to $\delta m_{e(2)}^2$ around the decoupling temperature and the average momentum of electron
\cite{Lopez:1998vk}, and contributes about $-0.00005$ to $N_{\rm eff}$ in the instantaneous decoupling limit \cite{Bennett:2019ewm}. Due to this smallness, we neglect the last term and consider only
the first two terms, which depend only on $T_{\gamma}$. On the other
hand, the thermal corrections to the photon mass at $\mathcal{O}(e^2)$
are given by \cite{Fornengo:1997wa},
\begin{align}
 \delta m_{\gamma(2)}^2(T_{\gamma}) = \frac{8\alpha}{\pi}\int^{\infty}_0dk\frac{k^2}{E_k}\frac{1}{\exp(E_k/T_{\gamma})+1}.
\end{align}
The total pressure and the total energy density of the electromagnetic
plasma are given by, including thermal mass corrections of electrons,
positrons and photons,
\begin{align}
 P &= \frac{T_{\gamma}}{\pi^2}\int^{\infty}_0dk~k^2\log \left[\frac{(1+e^{-E_e/T_{\gamma}})^2}{(1-e^{-E_\gamma/T_{\gamma}})} \right], \nonumber \\
 \rho &= -P+T_{\gamma}\frac{dP}{dT_{\gamma}},
\end{align}
where $E_{\gamma} =\sqrt{k^2 + \delta m_{\gamma}^2}$ and $E_e = \sqrt{k^2 + m_e^2 +
\delta m_e^2}$. $\delta m_{\gamma}^2$ and $\delta m_e^2$ denote the
thermal mass corrections of photons and electrons, respectively. We
expand $P$ in terms of $\delta m_e^2$ and $\delta m_{\gamma}^2$ at
$\mathcal{O}(e^2)$ and get the leading order correction to the pressure
\cite{Mangano:2001iu},
\begin{align}
 P_{(2)} = -\int^{\infty}_0 \frac{dk}{2\pi^2} \left[\frac{k^2}{E_k}\frac{\delta m_{e(2)}^2}{\exp(E_k/T_{\gamma})+1} +  \frac{k}{2}\frac{\delta m_{\gamma(2)}^2}{\exp(k/T_{\gamma})-1}  \right].
 \label{P_2}
\end{align} 
Here we need to introduce the symmetry factor $1/2$ in Eq.~(\ref{P_2})
in order to avoid the double counting of the thermal corrections to the
total pressure. Then the leading order correction to the energy density
is obtained as
\begin{align}
\rho_{(2)}=-P_{(2)}+T_{\gamma}\frac{dP_{(2)}}{dT_{\gamma}}.
\end{align}
 
The next-to-leading order of thermal corrections to the photon mass is
$\mathcal{O}(e^3)$. These nontrivial corrections to the photon mass
come from the resummation of ring diagrams at all orders. Through this
mass correction, the thermal corrections to the pressure and energy density are given by
\cite{Bennett:2019ewm},
\begin{align}
 P_{(3)} = \frac{e^3T_{\gamma}}{12\pi^4}I^{3/2}(T_{\gamma}), \nonumber \\
 \rho_{(3)} = \frac{e^3T_{\gamma}^2}{8\pi^4}I^{1/2}\partial_{T_{\gamma}}I,
\end{align} 
where
\begin{align}
I(T_{\gamma}) = 2\int^{\infty}_0 dk \left( \frac{k^2+E_k^2}{E_k} \right)\frac{1}{\exp(E_k/T_{\gamma})+1}.
\end{align}
Note that the thermal corrections at $\mathcal{O}(e^3)$ do not
modify the collision terms since these corrections change only the
photon mass while the next corrections to the electron mass would appear
at $\mathcal{O}(e^4)$.
Finally, we read the total energy density and the total pressure of the electromagnetic plasma  as
\begin{align}
 P &= P_{(0)}+P_{(2)}+P_{(3)}, \nonumber \\
 \rho &= \rho_{(0)} + \rho_{(2)}+\rho_{(3)}. 
\end{align}

 %%%%%%%%%%%%%%%%%%%%%%%%%%%%%%%%%%%%%%%%%%%%%%%%%%%%%%%%%%%%%%%%%%%%%%%%%%%%%%%%%%%%%%%%%%%%%%%%%%
 \subsection{Computational method and initial conditions}
 
We solve a set of Eqs.~(\ref{BE}) and (\ref{EC}) with the following
comoving variables instead of the cosmic time $t$, a momentum $p$, and
the photon temperature $T_{\gamma}$,
\begin{align}
x = m_ea,\ \ \ \ \ \ \  y = pa,\ \ \ \ \ \ \ \  z = T_{\gamma}a, 
\end{align}
where we take an arbitrary mass scale to be the electron mass $m_e$ and
$a$ is the scale factor of the Universe, normalized as $z \rightarrow 1$
in the high temperature limit. 
%$a(t)= 1/T_{\gamma}$ at very high temperature.

Since the Boltzmann equations (\ref{BE}) are integro-differential
equations due to the collision terms, these equations were solved by a
discretization in a grid of comoving momenta $y_i$ in
refs.~\cite{Hannestad:1995rs, Dolgov:1997mb, Dolgov:1998sf,
Grohs:2015tfy, Mangano:2005cc, deSalas:2016ztq, Gariazzo:2019gyi}, by an
expansion of the distortions of neutrinos in moments in
refs.~\cite{Esposito:2000hi, Mangano:2001iu, Birrell:2014uka}, or by a hybrid method combining the former two methods in ref.~\cite{Froustey:2019owm}. In this
study, we use the former discretization method and take 100 grid points
for the comoving momentum, equally spaced in the region $y_i \in [0.02,\
20]$.

We have numerically calculated the evolution of the density matrix and
the photon temperature in the interval $x_{\rm in}\leq x \leq x_{f}$. We
have chosen $x_{\rm in}=m_e/10\ {\rm MeV}$ as an initial time. Since
neutrinos keep in thermal equilibrium with the electromagnetic plasma at
$x_{\rm in}$, the initial values of density matrix $\rho_{y_i}^{\rm
in}(x)$ are,
\begin{align}
\rho_{y_i}^{\rm in}(x) = {\rm diag}\left(\frac{1}{e^{y_i/z_{\rm in}}+1},\frac{1}{e^{y_i/z_{\rm in}}+1},\frac{1}{e^{y_i/z_{\rm in}}+1}  \right).
\end{align} 
The initial value of the dimensionless photon temperature at $x_{\rm
in}$, $z_{\rm in}$, slightly differs from unity because of the finite
electron and positron masses.  Due to the entropy conservation of
electromagnetic plasma, neutrinos and anti-neutrinos, $z_{\rm in}$ is
set as in \cite{Dolgov:1998sf},
\begin{align}
z_{\rm in} = 1.00003.
\end{align}
We set $x_f = 30$ as a final time, when the neutrino density matrix and
$z$ can be regarded as frozen.

%%%%%%%%%%%%%%%%%%%%%%%%%%%%%%%%%%%%%%%%%%%%%%%%%%%%%%%%%%%%%%%%%%%%%%%%%%%%%%%%%%%%%%%%%%%%%%%%%%
%%%%%%%%%%%%%%%%%%%%%%%%%%%%%%%%%%%%%%%%%%%%%%%%%%%%%%%%%%%%%%%%%%%%%%%%%%%%%%%%%%%%%%%%%%%%%%%%%%
%%%%%%%%%%%%%%%%%%%%%%%%%%%%%%%%%%%%%%%%%%%%%%%%%%%%%%%%%%%%%%%%%%%%%%%%%%%%%%%%%%%%%%%%%%%%%%%%%%

\section{Results}
\label{sec:3}

\subsection{The flavor basis}

First, we have numerically solved a set of Eqs.~(\ref{BE}) and
(\ref{EC}) in the flavor basis, during the process of neutrino
decoupling. In order to compare with previous results, we discuss the
cases with and without neutrino mixing, and those with and without
finite temperature corrections from QED up to $\mathcal{O}(e^2)$ and
$\mathcal{O}(e^3)$. In the case with neutrino mixing, we also consider
the normal neutrino mass hierarchy with the latest best-fit values of
neutrino mixing parameters. Although we have considered the inverted
mass hierarchy too, the results are almost the same in the case of
normal mass hierarchy as in ref.~\cite{deSalas:2016ztq}. Hereafter we
only show the case of the normal mass hierarchy.

\begin{figure}[htb]
   \begin{center}
     \includegraphics[clip,width=11.8cm]{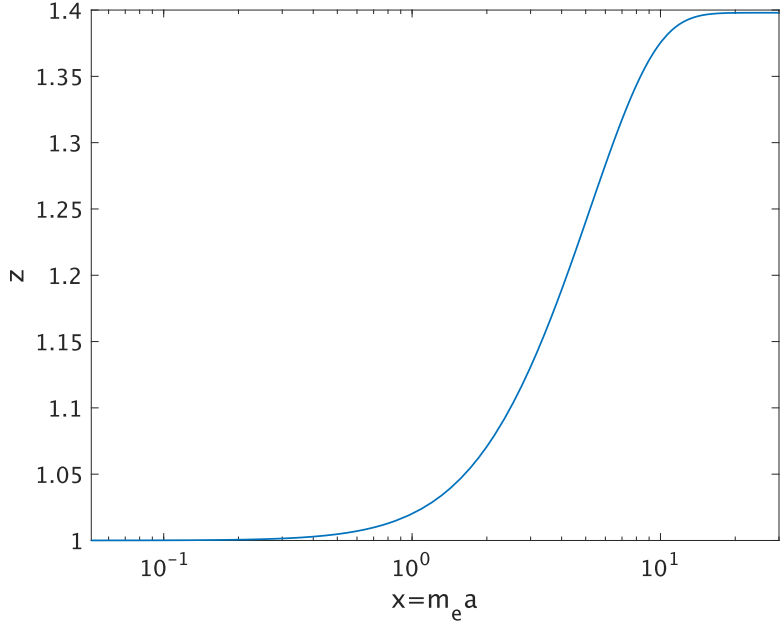}
    \end{center}
    \vspace{-8mm}
 \caption{The time evolution of the comoving photon temperature $z(x)$
 as a function of the normalized scale factor $x=m_ea$ in the case both with neutrino mixing and QED finite temperature corrections up to $\mathcal{O}(e^3)$
 .}
 \label{fig:x-z}
 \end{figure}

\begin{figure}[h]
   \begin{center}
     \includegraphics[clip,width=11.8cm]{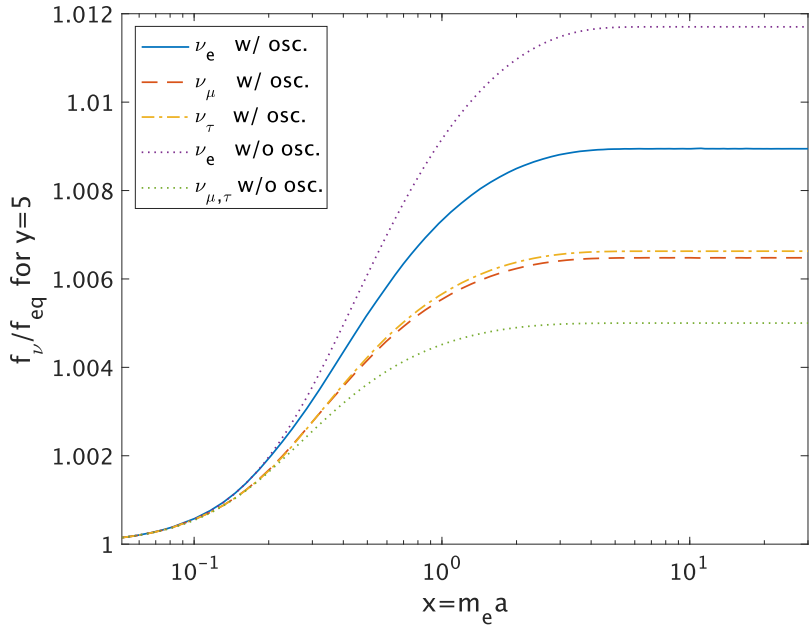}
    \end{center}
    \vspace{-8mm}
 \caption{The time evolution of the distortions of flavor neutrinos for a
 fixed momentum ($y=5$) as a function of the normalized scale factor
 $x=m_ea$ in the case with QED finite temperature corrections up to $\mathcal{O}(e^3)$.  Upper (lower) dotted line is for $\nu_e\ (\nu_{\mu,\tau})$
 without neutrino oscillations. Inner three lines are for flavor
 neutrinos with neutrino oscillations.}  \label{fig:flavorx-f}
 \end{figure}

In Figs.~\ref{fig:x-z} and \ref{fig:flavorx-f}, we show the evolution of
the photon temperature and the distortions of the flavor neutrino spectra
for a comoving momentum $(y=5)$ in the case with QED corrections up to $\mathcal{O}(e^3)$ respectively, where we plot the comoving
photon temperature $z(x)$ and the neutrino spectra $f_{\nu_\alpha}/f_{\rm eq}\ (f_{\rm eq} =
[\exp(y)+1]^{-1})$ as a function of the normalized cosmic scale factor
$x$. At high temperature with $(x \lesssim 0.2)$, the temperature
differences between photons and neutrinos are negligible and neutrinos
are in thermal equilibrium with electrons and positrons. In the
intermediate regime with $(0.2 \lesssim x \lesssim 4)$, weak interactions
gradually become ineffective with shifting from small to large
momenta. In this period, the neutrino spectra are distorted since
the energies of electrons and positrons partially convert into those of
neutrinos coupled with electromagnetic plasma. Finally, at
low temperature with $(x > 4)$, the collision term $C[\rho_p(t)]$
becomes ineffective and the distortions are frozen.

In Fig.~\ref{fig:flavorx-f}, we show the results of the two cases
with and without neutrino mixing. We find that the final values of
$f_{\nu_\alpha}[y=5]$ without flavor mixing are $1.17\%$ for $\nu_e$
and $0.500\%$ for $\nu_{\mu,\tau}$ larger than those in
the instantaneous decoupling limit, that is, $f_{\rm eq}[y=5]$. This
difference between electron-type neutrinos and mu(tau)-type neutrinos
arises from the fact that only electron-type neutrinos interact with
electrons and positrons through the weak charged-currents. On the other
hand, in the cases with neutrino mixing, neutrino oscillations
mix the distortions of the flavor neutrinos too. Though the flavor
oscillation effects are subdominant to the refractive effects in the period
with $(x \lesssim 0.2)$, the refractive term gets ineffective in the lower
temperature due to the annihilations of electrons and positrons. Thus,
the oscillation terms finally mix the flavor neutrino distortions. We
also find that the final values of $f_{\nu_\alpha}[y=5]$ with flavor mixing
are $0.895\%$ for $\nu_e$, $0.648\%$ for $\nu_{\mu}$ 
and $0.663\%$ for $\nu_{\tau}$ larger than those in the
instantaneous decoupling limit, that is, $f_{\rm eq}[y=5]$.

\begin{figure}[htb]
   \begin{center}
     \includegraphics[clip,width=11.8cm]{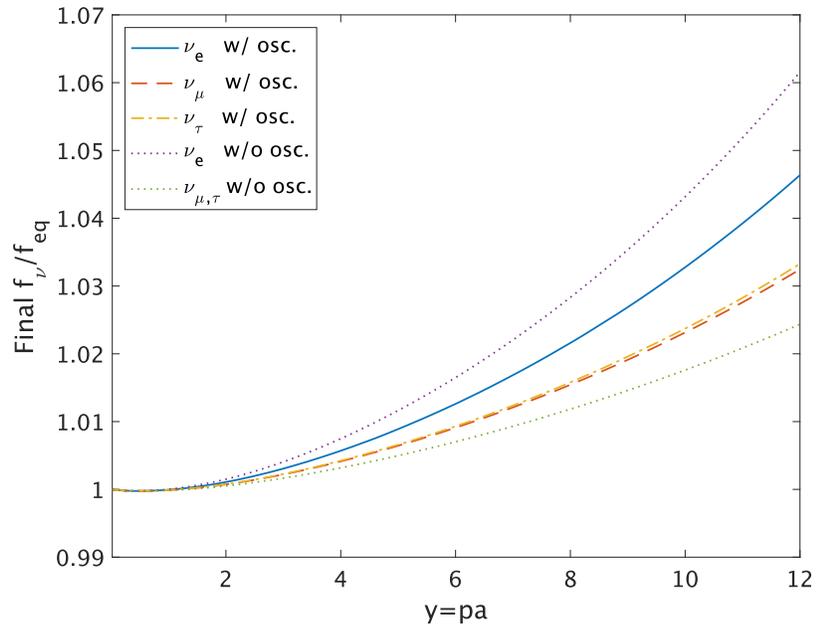}
    \end{center}
    \vspace{-8mm}
 \caption{The final distortions of flavor neutrino spectra as a function
 of the comoving momentum $y$ in the case with QED finite temperature corrections up to $\mathcal{O}(e^3)$. Dotted lines represent those for
 $\nu_e\ (\nu_{\mu,\tau})$ without neutrino oscillations, while solid and
 dashed lines represent those for flavor neutrinos with neutrino
 oscillations.}  \label{fig:flavory-f}
 \end{figure}

In Fig.~\ref{fig:flavory-f}, we show the frozen values of the flavor
neutrino spectra $f_{\nu_\alpha}/f_{\rm eq}$ as a function of a comoving
momentum $y$ for both cases with and without neutrino mixing, including
QED corrections up to $\mathcal{O}(e^3)$. This figure shows the fact
that neutrinos with higher energies interact with electrons and
positrons until a later epoch. In addition, we see that neutrino
oscillations tend to equilibrate the flavor neutrino
distortions. Although the neutrino spectra $f_{\nu_\alpha}/f_{\rm eq}$
with low energies are very slightly less than unity, these
extractions of low energy neutrinos stem from an energy boost through
the scattering by electrons, positrons, (and neutrinos) with
sufficiently high energies, which are not yet annihilated and hence
still effective at the neutrino decoupling process.

We also give several important quantities charactering the
decoupling process of neutrinos. In Tables.~\ref{tb:flavor} and
\ref{tb:flavor2}, we give final values (at $x_f = 30$) of the
dimensionless photon temperature $z_{\rm fin}$, the difference of energy
densities and number densities of flavor neutrinos from those in the
instantaneous decoupling limit denoted by $\rho_{\nu_0}$ and
$n_{\nu_0}$, and the effective number of neutrinos $N_{\rm eff}$ defined as
\begin{align}
\rho_r = \left[ 1 + \frac{7}{8}\left(\frac{4}{11}\right)^{4/3}N_{\rm eff}\right]\rho_{\gamma}.
\end{align}
Here $\rho_r$ and $\rho_{\gamma}$ are the energy densities of the total
radiations and photons, respectively. The effective number of neutrinos
$N_{\rm eff}$ can be rewritten,
\begin{align}
N_{\rm eff} = \left( \frac{z_0}{z_{\rm fin}} \right)^4 \left(3+ \frac{\delta \rho_{\nu_e}}{\rho_{\nu_0}}+ \frac{\delta \rho_{\nu_\mu}}{\rho_{\nu_0}}+ \frac{\delta \rho_{\nu_\tau}}{\rho_{\nu_0}}\right),
\label{Neff}
\end{align}
where $z_0 = (11/4)^{1/3}\simeq 1.40102$ is the final value of the dimensionless
photon temperature in the instantaneous decoupling limit and $\delta
\rho_{\nu_\alpha}=\rho_{\nu_\alpha}-\rho_{\nu_0}$.

Without neutrino mixing, we find the final values of $N_{\rm eff}$
are 3.03404 for the case without QED corrections and
3.04430 for the case with those up to $\mathcal{O}(e^2)$, which agree
with recent previous works~\cite{Birrell:2014uka, Grohs:2017iit,
Froustey:2019owm}. In addition, our results for the case without
neutrino mixing but with QED corrections up to
$\mathcal{O}(e^3)$ show that the final value of $N_{\rm eff}$ is
slightly modified to 3.04335. Thus, the difference with QED corrections up to $\mathcal{O}(e^2)$ and
$\mathcal{O}(e^3)$ (but without neutrino mixing) is 0.00095 in terms of
$N_{\rm eff}$, which is very close to the value estimated in the instantaneous decoupling limit \cite{Bennett:2019ewm}.

In the cases with neutrino mixing,
Table.~\ref{tb:flavor2} shows that the energy densities of
$\mu,\tau$-type neutrinos increase more while those of
electron-type neutrinos increase less, compared to the cases without neutrino mixing. This modification leads to the enhancement of the total
energy density for neutrinos with final values of $N_{\rm eff}=3.04486$
with QED corrections up to $\mathcal{O}(e^2)$ and $N_{\rm eff}=3.04391$
with QED corrections up to $\mathcal{O}(e^3)$. 
Since the blocking factor for electron neutrinos, $(1-f_{\nu_e}),$ is decreased
 by neutrino mixing, the annihilation of electrons and positrons into electron neutrinos increases. 
 Although the annihilation into the other neutrinos decreases, electron neutrinos contribute to the neutrino heating most efficiently, and
 neutrino oscillations enhance the annihilation of electrons and positrons into neutrinos.
From these processes, we
conclude that neutrino oscillations slightly promote neutrino
heating and the difference of $N_{\rm eff}$ is $0.00056$, which
agrees with the results of previous
works\cite{deSalas:2016ztq,Hannestad:2001iy,Escudero:2020dfa}. In this
case, we also find that the difference of $N_{\rm eff}$ between the cases
 including QED corrections up to $\mathcal{O}(e^2)$ and $\mathcal{O}(e^3)$ is 0.00095.
 
 Finally, we comment on errors of the results due to the numerical calculations and the choice of physical parameters.
Our numerical calculations converge very well, which is confirmed by the same values of QED corrections at $\mathcal{O}(e^3)$ in the two cases with and without neutrino mixing.
However, the results could be affected by neglecting some QED corrections such as the second line in Eq.~(\ref{delta_me}), neglecting the off-diagonal terms for the self-interactions of neutrinos and dependence on the values of mixing parameters, which would change $N_{\rm eff}$
by less than $0.0005$. Taking into account these errors, we conclude $N_{\rm eff}=3.044$ for the most accurate case.

\vspace{+2cm}

\begin{table}[h]
\begin{center}
\small
  \begin{tabular}{|c|c|c|c|c|c|} \hline
    Case                                                              & $z_{\rm fin}$ & $N_{\rm eff}$ \\ \hline
    Instantaneous decoupling                                        & 1.40102        & 3.000  \\
    No mixing\hspace{-0.4mm} +\hspace{-0.4mm} No QED                                            & 1.39910        & 3.034  \\
    \hspace{-0.8mm}No mixing\hspace{-0.4mm} +\hspace{-0.4mm} QED up to $\mathcal{O}(e^2)$\hspace{-0.8mm}        & 1.39789        & 3.044  \\ 
     \hspace{-0.8mm}No mixing\hspace{-0.4mm} +\hspace{-0.4mm} QED up to $\mathcal{O}(e^3)$\hspace{-0.8mm}        & 1.39800        & 3.043  \\ \hline
    mixing\hspace{-0.4mm} +\hspace{-0.4mm} QED up to $\mathcal{O}(e^2)$              & 1.39786        & 3.045  \\
    mixing\hspace{-0.4mm} +\hspace{-0.4mm} QED up to $\mathcal{O}(e^3)$              & 1.39797        & 3.044 \\ \hline
  \end{tabular}
  \caption{The final values of comoving photon temperature and the effective number of neutrinos for flavor neutrinos in several cases.}
  \label{tb:flavor}
  \end{center}
\end{table}

\vspace{+1cm}

\begin{table}[h]
\begin{center}
\small
  \begin{tabular}{|c|c|c|c|c|c|c|} \hline
    Case                                                              & $\delta \bar{\rho}_{\nu_e}(\%)$ & $\delta \bar{\rho}_{\nu_\mu}(\%)$ & $\delta \bar{\rho}_{\nu_\tau}(\%)$ & $\delta \bar{n}_{\nu_e}(\%)$ & $\delta \bar{n}_{\nu_\mu}(\%)$ & $\delta \bar{n}_{\nu_\tau}(\%)$ \\ \hline
    Instantaneous decoupling                                               &0   &0  &0  & 0  & 0  & 0\\
    No mixing\hspace{-0.4mm} +\hspace{-0.4mm} No QED                                                   &0.949   &0.397  &0.397       & 0.583 & 0.240 & 0.240  \\
    \hspace{-0.8mm}No mixing\hspace{-0.4mm} +\hspace{-0.4mm} QED up to $\mathcal{O}(e^2)$\hspace{-0.8mm}               &0.937   &0.391  &0.391       & 0.575 & 0.236 & 0.236 \\ 
    \hspace{-0.8mm}No mixing\hspace{-0.4mm} +\hspace{-0.4mm} QED up to $\mathcal{O}(e^3)$\hspace{-0.8mm}               &0.937   &0.391  &0.391       & 0.575 & 0.236 & 0.236   \\ \hline
    mixing\hspace{-0.4mm} +\hspace{-0.4mm} QED up to $\mathcal{O}(e^2)$                     &0.712   &0.511  &0.523       & 0.435 & 0.311 & 0.319   \\
    mixing\hspace{-0.4mm} +\hspace{-0.4mm} QED up to $\mathcal{O}(e^3)$                     &0.712   &0.511   &0.523      & 0.436 & 0.312 & 0.319\\ \hline
  \end{tabular}
  \caption{The final values of the distortions of energy densities
  $\delta \bar{\rho}_{\nu_\alpha} \equiv \delta
  \rho_{\nu_\alpha}/\rho_{\nu_0}$ and number densities $\delta
  \bar{n}_{\nu_\alpha} \equiv (n_{\nu_\alpha}-n_{\nu_0})/n_{\nu_0}$
  for flavor neutrinos in several cases.}  \label{tb:flavor2}
  \end{center}
\end{table}

\clearpage
%%%%%%%%%%%%%%%%%%%%%%%%%%%%%%%%%%%%%%%%%%%%%%%%%%%%%%%%%%%%%%%%%%%%%%%%%%%%%%
\subsection{The mass basis}

In this section, we present the results of numerical calculations with a
set of Eqs.~(\ref{EC}) and (\ref{BE2}). In the mass basis, we have also
solved these equations with and without QED finite temperature corrections
up to $\mathcal{O}(e^2)$ and $\mathcal{O}(e^3)$. Also in the mass basis,
we show only the case of the normal mass hierarchy with the best-fit values of
mixing parameters of neutrinos since we have checked that the results in the normal and
inverted hierarchies are almost the same.

\begin{figure}[htbp]
   \begin{center}
     \includegraphics[clip,width=11.8cm]{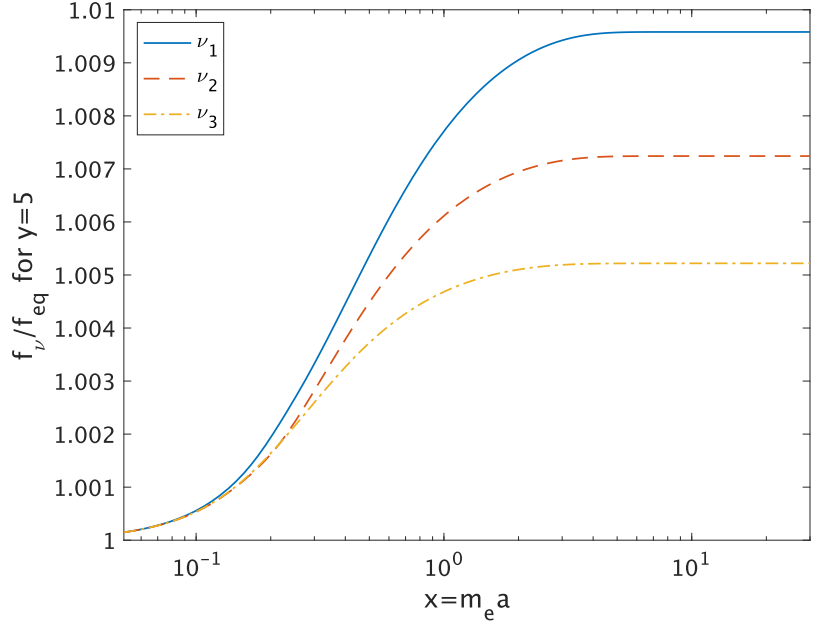}
    \end{center}
    \vspace{-8mm}
 \caption{The time evolution of the distortions of massive neutrinos for a fixed momentum ($y=5$) in the case with QED finite temperature corrections up to $\mathcal{O}(e^3)$.}
 \label{fig:massx-f}
 \end{figure}

\begin{figure}[htbp]
   \begin{center}
     \includegraphics[clip,width=11.8cm]{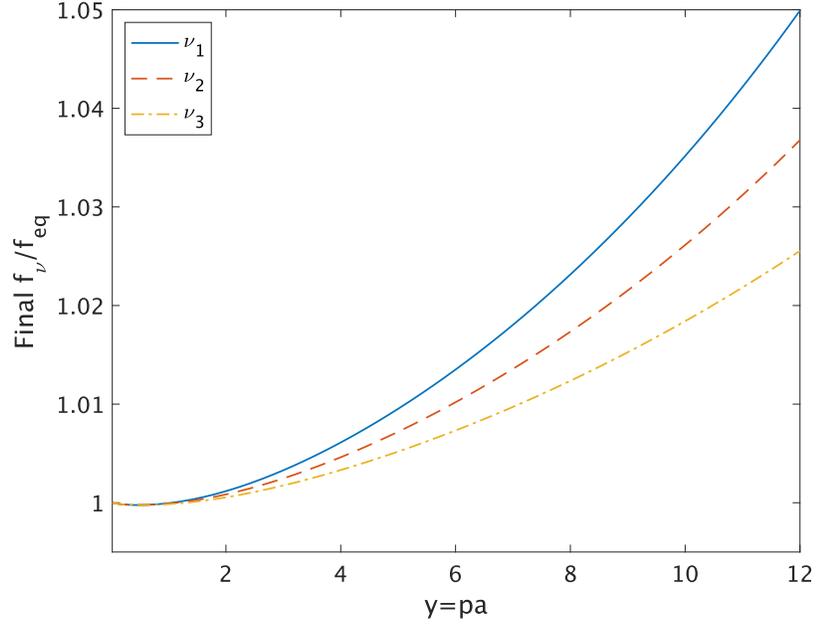}
    \end{center}
    \vspace{-8mm}
 \caption{The final distortions of massive neutrino spectra as a function of the comoving momentum $y$ in the case with QED finite temperature corrections up to $\mathcal{O}(e^3)$.}
 \label{fig:massy-f}
 \end{figure}

In Fig.~\ref{fig:massx-f}, we show that the evolution of the massive
neutrino spectra, $f_{\nu_i}/f_{\rm eq}$, for a comoving momentum $(y=5)$
as a function of the normalized scale factor $x$ with QED corrections up to 
$\mathcal{O}(e^3)$. The final values of
$f_{\nu_i}[y=5]$ are found to be $0.958\%$ for $\nu_1$, $0.724\%$ for
$\nu_2$, and $0.522\%$ for $\nu_3$ larger than those in the
instantaneous decoupling limit, $f_{eq}[y=5]$. These differences of
distortions arise since each massive neutrino interacts with electrons
and positrons through the different coupling $Z^L$ in Eq.~(\ref{CM}) and
the refractive effects in the mass basis generate the off-diagonal parts
of the mass matrix for massive neutrinos effectively. In
Fig.~\ref{fig:massy-f}, we show the asymptotic values of the massive
neutrino spectra $f_{\nu_i}/f_{\rm eq}$ as a function of $y$ 
with QED corrections up to $\mathcal{O}(e^3)$.

In Tables.~\ref{tb:mass} and \ref{tb:mass2}, we also show that the final
values of the dimensionless photon temperature $z_{\rm fin}$, the energy
densities $\rho_{\nu_i}/\rho_{\nu_0}$ and number densities $n_{\nu_i}/n_{\nu_0}$ of massive neutrinos
, and the effective
number of neutrinos $N_{\rm eff}$. For the cases with finite
temperature corrections up to $\mathcal{O}(e^2)$ and $\mathcal{O}(e^3)$,
we find very good agreement for $z_{\rm fin}$ and $N_{\rm eff}$ both in
the mass basis and in the flavor basis (with neutrino mixing). The final
values of $N_{\rm eff}$ are 3.04483 for the case with QED corrections up
to $\mathcal{O}(e^2)$ and 3.04388 for the case with those up to
$\mathcal{O}(e^3)$. We also find that the difference of $N_{\rm eff}$
between the cases including QED corrections up to $\mathcal{O}(e^2)$ and
$\mathcal{O}(e^3)$ is 0.00095, which is the same in the
flavor basis. The small difference for $N_{\rm eff}$ in both bases may
come from the fact that we neglect the off-diagonal parts for
self-interaction processes in the collision terms of the Boltzmann
equations.

\begin{table}[h]
\begin{center}
\small
  \begin{tabular}{|c|c|c|c|c|c|} \hline
    Case                                                              & $z_{\rm fin}$ &  $N_{\rm eff}$ \\ \hline
     QED up to $\mathcal{O}(e^2)$        & 1.39786              & 3.045  \\
     QED up to $\mathcal{O}(e^3)$        & 1.39797              & 3.044  \\ \hline
  \end{tabular}
  \caption{The final values of comoving photon temperature and the effective number of neutrinos for massive neutrinos in several cases.}
  \label{tb:mass}
  \end{center}
\end{table}

\begin{table}[h]
\begin{center}
\small
  \begin{tabular}{|c|c|c|c|c|c|c|} \hline
    Case                                                              & $\delta \bar{\rho}_{\nu_1}(\%)$ & $\delta \bar{\rho}_{\nu_2}(\%)$ & $\delta \bar{\rho}_{\nu_3}(\%)$ & $\delta \bar{n}_{\nu_1}(\%)$ & $\delta \bar{n}_{\nu_2}(\%)$ & $\delta \bar{n}_{\nu_3}(\%)$ \\ \hline
   QED up to $\mathcal{O}(e^2)$                     &0.764   &0.573  &0.409       & 0.468 & 0.350 & 0.248   \\
   QED up to $\mathcal{O}(e^3)$                     &0.764   &0.574   &0.409      & 0.468 & 0.350 & 0.248\\ \hline
  \end{tabular}
  \caption{The final values of the distortions of energy densities $\delta \bar{\rho}_{\nu_i} \equiv \delta \rho_{\nu_i}/\rho_{\nu_0}$ and number densities $\delta \bar{n}_{\nu_i} \equiv  (n_{\nu_i}-n_{\nu_0})/n_{\nu_0}$ for massive neutrinos in several cases.}
  \label{tb:mass2}
  \end{center}
\end{table}

%%%%%%%%%%%%%%%%%%%%%%%%%%%%%%%%%%%%%%%%%%%%%%%%%%%%%%%%%%%%%%%%%%%%%%%%%%%%%
\subsection{Transformation of distributions in the flavor and mass bases}

In this section, we derive the relation between the distribution functions
in the flavor and mass bases in ultra-relativistic limit and check
the numerical results given in previous sections. The relation of
annihilation operators for negative-helicity neutrinos between flavor and mass eigenstates
is given by,
\begin{align}
a_\alpha(\bm{p},t) = \sum_{i=1,2,3} U_{\alpha i}a_i(\bm{p},t),
\end{align}
with $\alpha=e,\mu,\tau$. Using the above relation, we describe the density operators for flavor neutrinos $a_\beta^{\dag}a_\alpha$ through the operators for massive neutrinos,
\begin{align}
a^{\dag}_\beta(\bm{p},t) a_\alpha(\bm{p}',t) = \sum_{i,j=1,2,3} U^{\ast}_{\beta j}U_{\alpha i}a^{\dag}_j(\bm{p},t)a_i(\bm{p}',t),
\end{align}
 and we get the distribution functions for flavor neutrinos as the density matrix for massive neutrinos easily,
 \begin{align}
 f_{\nu_\alpha}(\bm{p},t) = \sum_{i,j=1,2,3} U^{\ast}_{\alpha j}U_{\alpha i}(\rho_p)_{ij}.
 \end{align}
In particular, after the decoupling  process of neutrinos, the off-diagonal parts of neutrino density matrix in the mass basis are expected to vanish since all interactions involving neutrinos are ineffective in this period and neutrinos in the mass basis do not oscillate. Then the relation between distribution functions in the flavor and mass bases after the decoupling of neutrinos is given by,
\begin{align}
f_{\nu_\alpha}(\bm{p},t) = \sum_i |U_{\alpha i}|^2f_{\nu_i}(\bm{p},t).
\label{relation}
\end{align}
We have numerically confirmed Eq.(\ref{relation}) and $(\rho_p(t))_{ij}\simeq0\ (i \neq j)$ after the decoupling of neutrinos.
Using Eq.~(\ref{relation}) and Tables in the mass basis, we can estimate
$N_{\rm eff}=3.04389$ in the flavor basis, which is very close to the
numerical value of $N_{\rm eff}=3.04391$ in the flavor basis. 

%%%%%%%%%%%%%%%%%%%%%%%%%%%%%%%%%%%%%%%%%%%%%%%%%%%%%%%%%%%%%%%%%%%%%%%%%%%%%%
%%%%%%%%%%%%%%%%%%%%%%%%%%%%%%%%%%%%%%%%%%%%%%%%%%%%%%%%%%%%%%%%%%%%%%%%%%%%%%
%%%%%%%%%%%%%%%%%%%%%%%%%%%%%%%%%%%%%%%%%%%%%%%%%%%%%%%%%%%%%%%%%%%%%%%%%%%%%%
\section{Conclusions}
\label{sec:con}

We have studied the neutrino decoupling process in the early Universe by
solving the kinetic equations for neutrinos numerically. We have
calculated the evolution of the neutrino spectral distortions not only
in the flavor basis but also in the mass basis. The latter approach
enables us to easily reveal the neutrino momentum spectra at the current
Universe in future work. The calculations in both bases are also
useful for the cross-check of the results.  In addition, in preparation
for precision measurements of the effective number of neutrinos $N_{\rm
eff}$, we have also considered the effects due to QED finite temperature
corrections up to $\mathcal{O}(e^3)$ on the relevant kinetic equations
for the first time.

In both bases, we have solved the momentum-dependent kinetic
equations for the neutrino density matrix, where we have considered the
full collision terms for the processes including neutrinos, electrons
and their anti-particles while we neglect the off-diagonal parts of the
collision terms for neutrino self-interaction processes. We find in
both bases that the effective number of neutrinos is $N_{\rm eff} =
3.044$. The effects of neutrino oscillations increase $N_{\rm eff}$ by
about 0.0005 compared to $N_{\rm eff}$ without neutrino oscillations
since neutrino oscillations promote the annihilation of
electron-positron pairs into neutrinos.  On the other hand, the impacts
of QED corrections up to $\mathcal{O}(e^3)$ decrease $N_{\rm eff}$ by
about 0.001 compared to $N_{\rm eff}$ with QED corrections up to
$\mathcal{O}(e^2)$. 
The estimated error for the value of $N_{\rm eff}$ would be at most
$0.0005$, which mainly stems from neglecting some finite temperature
corrections and the off-diagonal terms for self-interactions of neutrinos,
and from the dependence on the values of mixing parameters.
We also find the final values of the number density
and energy density for each neutrino. In particular, these values in the
mass basis may be important for detection processes of relic neutrinos
in the current Universe.

The current constraint on $N_{\rm eff}$ from the Planck data analyses \cite{Akrami:2018vks}
in $\Lambda$CDM model is $N_{\rm eff} = 2.99^{+0.34}_{-0.33}$ at $95\%$ CL \cite{Aghanim:2018eyx}, 
consistent with our prediction, $N_{\rm eff}=3.044$.
Upcoming CMB and LSS experiments are expected to improve neutrino masses 
and energy density bounds over the next years (see e.g. \cite{Benson:2014qhw, Ade:2018sbj}) and determine $N_{\rm eff}$ with $1\%$ precision in the near future (see e.g. \cite{Abazajian:2013oma, DiValentino:2016foa, Hanany:2019lle, Sehgal:2019ewc, Abazajian:2019eic}).

Finally, in ultra-relativistic limit and after the decoupling for
neutrinos, we also find the simple transformation formula between the
distribution functions in the flavor and mass bases, which is confirmed
by our numerical calculation. Using this relation, we can easily switch
the distribution functions in the flavor and mass bases for neutrinos
without direct numerical calculations in the two bases.

%%%%%%%%%%%%%%%%%%%%%%%%%%%%%%%%%%%%%%%%%%%%%%%%%%%%%%%%%%%%%%%%%%%%%%%%%%%%%%
%%%%%%%%%%%%%%%%%%%%%%%%%%%%%%%%%%%%%%%%%%%%%%%%%%%%%%%%%%%%%%%%%%%%%%%%%%%%%%
%%%%%%%%%%%%%%%%%%%%%%%%%%%%%%%%%%%%%%%%%%%%%%%%%%%%%%%%%%%%%%%%%%%%%%%%%%%%%%

\section*{Acknowledgments}

We are grateful to Shoichi Yamada for useful discussions.
KA is supported by JSPS Grant-in-Aid for Research Fellows
No. 19J14449. KA and MY are supported in part by JSPS Bilateral Open
Partnership Joint Research Projects. MY is supported in part by JSPS
Grant-in-Aid for Scientific Research Numbers 18K18764 and Mitsubishi
Foundation.

%%%%%%%%%%%%%%%%%%%%%%%%%%%%%%%%%%%%%%%%%%%%%%%%%%%%%%%%%%%%%%%%%%%%%%%%%%%%%%
 %%%%%%%%%%%%%%%%%%%%%%%%%%%%%%%%%%%%%%%%%%%%%%%%%%%%%%%%%%%%%%%%%%%%%%%%%%%%%%
%%%%%%%%%%%%%%%%%%%%%%%%%%%%%%%%%%%%%%%%%%%%%%%%%%%%%%%%%%%%%%%%%%%%%%%%%%%%%%
\appendix

\section{Kinetic equations for neutrinos in comoving variables}
\label{appa}
In this appendix, we write the Boltzmann equations and the energy conservation law in terms of the comoving variables, $x=m_ea,\ y=pa,\ z=T_{\gamma}a $. In terms of these variables, we can write the Boltzmann equations (\ref{BE}) as in \cite{deSalas:2016ztq},
\begin{align}
\frac{d\rho_y(x)}{dx}=m_{Pl}\sqrt{\frac{3}{8\pi\bar{\rho}}}\left\{-i\frac{x^2}{m_e^3}\left[\left(\frac{M^2}{2y}-\frac{8\sqrt{2}G_Fym_e^6}{3m_W^2x^6}\bar{E} \right), \rho_y(x)\right]+\frac{m_e^3}{x^4}\bar{C}[\rho_y(x)] \right\}.
\end{align}
where $\bar{\rho}$, $\bar{E}$, and $\bar{C}[\rho_y(x)]$ are quantities written in the comoving variables, $x,\ y,\ z$. We can write $\bar{\rho}$ and $\bar{E}$ as 
\begin{align}
\bar{\rho}&=\rho\left(\frac{x}{m_e}\right)^4, \nonumber \\
\bar{E}&={\rm diag}\left(\rho_{ee}\left(\frac{x}{m_e}\right)^4 ,0,0\right).
\end{align}
We also give the diagonal collision term from the self-interaction processes in the comoving variables $\bar{C}_S[\nu_\alpha(y_1)]$, where nine-dimensional collision integrals are reduced to two-dimensional collision integrals as in appendix~\ref{appb},
\begin{align}
\bar{C}_S[\nu_\alpha(y_1)]&=\frac{G_F^2}{2\pi^3y_1}\int dy_2dy_3\ y_2y_3y_4 \biggl[\Pi_S^1F(\nu_{\alpha}^{(1)},\nu_{\alpha}^{(2)},\nu_{\alpha}^{(3)},\nu_{\alpha}^{(4)}) \nonumber \\
&\ \ \ \ \ \ \ \ \ \ \ \  +\Pi_S^2F(\nu_{\alpha}^{(1)},\nu_{\beta}^{(2)},\nu_{\alpha}^{(3)},\nu_{\beta}^{(4)}) +\Pi_S^3F(\nu_{\alpha}^{(1)},\nu_{\alpha}^{(2)},\nu_{\beta}^{(3)},\nu_{\beta}^{(4)}) \nonumber \\
&\ \ \ \ \ \ \ \ \ \ \ \  +\Pi_S^2F(\nu_{\alpha}^{(1)},\nu_{\gamma}^{(2)},\nu_{\alpha}^{(3)},\nu_{\gamma}^{(4)}) +\Pi_S^3F(\nu_{\alpha}^{(1)},\nu_{\alpha}^{(2)},\nu_{\gamma}^{(3)},\nu_{\gamma}^{(4)}) \biggl].
\label{CSC}
\end{align}
Similarly, the collision terms from the annihilation processes and scattering processes are
\begin{align}
\bar{C}_A&=\frac{G_F^2}{2\pi^3y_1}\int dy_2dy_3\ y_2y_3\bar{E}_4 \nonumber \\
&\ \ \ \   \times \biggl[\Pi_A^1F^{LL}_A\left(\nu^{(1)}, \bar{\nu}^{(2)}, e^{(3)}, \bar{e}^{(4)}\right) +\Pi_A^2F^{RR}_A\left(\nu^{(1)}, \bar{\nu}^{(2)}, e^{(3)}, \bar{e}^{(4)}\right) \nonumber \\
&\ \ \ \ \ \ \ \   +\Pi_A^3\Bigl(F^{RL}_A\left(\nu^{(1)},\bar{\nu}^{(2)}, e^{(3)}, \bar{e}^{(4)} \right)+ F^{LR}_A\left(\nu^{(1)}, \bar{\nu}^{(2)}, e^{(3)}, \bar{e}^{(4)}\right)  \Bigl) \biggl], \label{CAC} \\
\bar{C}_{SC}&=\frac{G_F^2}{2\pi^3y_1}\int dy_2dy_3\ y_2y_3\bar{E}_4 \nonumber \\
&\ \ \ \   \times \biggl[\Pi_{SC}^1\Bigl(F^{LL}_{SC}\left(\nu^{(1)},e^{(2)}, \nu^{(3)},e^{(4)}\right) +F^{RR}_{SC}\left(\nu^{(1)},e^{(2)}, \nu^{(3)},e^{(4)}\right) \Bigl) \nonumber \\
&\ \ \ \ \ \ \ \ -\Pi_{SC}^2 \Bigl(F^{LR}_{SC}\left(\nu^{(1)},e^{(2)}, \nu^{(3)},e^{(4)}\right) +F^{RL}_{SC}\left(\nu^{(1)},e^{(2)}, \nu^{(3)},e^{(4)}\right) \Bigl) \biggl],
\label{CSCC}
\end{align}
where $E_i=\sqrt{y_i^2+x^2+\delta\bar{m}_e^2}$ and $\delta\bar{m}_e$ is the finite temperature correction to the electron mass up to $\mathcal{O}(e^2)$ in the comoving variables,
\begin{align}
\delta \bar{m}_e^2 = \frac{2\pi\alpha z}{3}+\frac{4\alpha}{\pi}\int dy \frac{y^2}{\sqrt{y^2+x^2}}\frac{1}{\exp(\sqrt{y^2+x^2}/z)+1}.
\end{align}
The functions $\Pi_{S,A,SC}^{1,2,3}$ in Eqs.~(\ref{CSC}), (\ref{CAC})
and (\ref{CSCC}) take the following forms,
\begin{align}
\Pi_S^1&=6D_1-\frac{4D_2(y_1,y_4)}{y_1y_4}-\frac{4D_2(y_2,y_3)}{y_2y_3}+\frac{2D_2(y_1,y_2)}{y_1y_2}+\frac{2D_2(y_3,y_4)}{y_3y_4}+\frac{6D_3}{y_1y_2y_3y_4}, \nonumber \\
\Pi_S^2&=2D_1+\frac{D_2(y_1,y_2)}{y_1y_2}+\frac{D_2(y_3,y_4)}{y_3y_4}-\frac{D_2(y_1,y_4)}{y_1y_4}-\frac{D_2(y_2,y_3)}{y_2y_3}+\frac{2D_3}{y_1y_2y_3y_4}, \nonumber \\
\Pi_S^3&=D_1-\frac{D_2(y_2,y_3)}{y_2y_3}-\frac{D_2(y_1,y_4)}{y_1y_4}+\frac{D_3}{y_1y_2y_3y_4}, \nonumber \\
\Pi_A^1&= 2D_1-\frac{2D_2(y_2,y_3)}{y_2\bar{E}_3}-\frac{2D_2(y_1,y_4)}{y_1\bar{E}_4}+\frac{2D_3}{y_1y_2\bar{E}_3\bar{E}_4}, \nonumber \\
\Pi_A^2&= 2D_1-\frac{2D_2(y_2,y_4)}{y_2\bar{E}_4}-\frac{2D_2(y_1,y_3)}{y_1\bar{E}_3}+\frac{D_3}{y_1y_2\bar{E}_3\bar{E}_4}, \nonumber \\
\Pi_A^3&= (x^2+\delta \bar{m}_e^2)\left(D_1+\frac{D_2(y_1,y_2)}{y_1y_2}\right)\frac{1}{\bar{E}_3\bar{E}_4}, \nonumber \\
\Pi_{SC}^1&=4D_1-\frac{2D_2(y_2,y_3)}{\bar{E}_2y_3}-\frac{2D_2(y_1,y_4)}{y_1\bar{E}_4}+\frac{2D_2(y_3,y_4)}{y_3\bar{E}_4}+\frac{2D_2(y_1,y_2)}{y_1\bar{E}_2}+\frac{4D_3}{y_1\bar{E}_2y_3\bar{E}_4}, \nonumber \\
\Pi_{SC}^2&= 2(x^2+\delta \bar{m}_e^2)\left(D_1-\frac{D_2(y_1,y_3)}{y_1y_3} \right)\frac{1}{\bar{E}_2\bar{E}_4}.
\end{align}
The functions of $D_{1,2,3}$ are written as,
\begin{align}
D_1&=\frac{4}{\pi}\int^{\infty}_0\frac{d\lambda}{\lambda^2} \sin(\lambda y_1) \sin(\lambda y_2) \sin(\lambda y_3) \sin(\lambda y_4), \nonumber \\
D_2(y_3,y_4)&=\frac{4y_3y_4}{\pi}\int^{\infty}_0\frac{d\lambda}{\lambda^2} \sin(\lambda y_1) \sin(\lambda y_2)\left[\cos(\lambda y_3)-\frac{\sin(\lambda y_3)}{\lambda y_3} \right]\left[\cos(\lambda y_4)-\frac{\sin(\lambda y_4)}{\lambda y_4} \right], \nonumber \\
D_3&=\frac{4y_1y_2y_3y_4}{\pi}\int^{\infty}_0\frac{d\lambda}{\lambda^2}\left[\cos(\lambda y_1)-\frac{\sin(\lambda y_1)}{\lambda y_1} \right]\left[\cos(\lambda y_2)-\frac{\sin(\lambda y_2)}{\lambda y_2} \right] \nonumber \\
&\ \ \ \ \ \ \ \ \ \ \ \ \ \ \ \ \ \ \ \  \times \left[\cos(\lambda y_3)-\frac{\sin(\lambda y_3)}{\lambda y_3} \right]\left[\cos(\lambda y_4)-\frac{\sin(\lambda y_4)}{\lambda y_4} \right],
\label{Dfunc}
\end{align}
which can be integrated analytically as in appendix~\ref{appb}.

Finally, the energy conservation law (\ref{EC}) is translated into the evolution equation for $z$, including the finite temperature corrections from QED up to $\mathcal{O}(e^3)$ \cite{Mangano:2001iu, Bennett:2019ewm},
\begin{align}
\frac{dz}{dx}=\frac{\frac{x}{z}J(x/z)-\frac{1}{2\pi^2z^3}\int^{\infty}_0dy\ y^3\left(\frac{df_{\nu_e}}{dx}+\frac{df_{\nu_\mu}}{dx}+\frac{df_{\nu_\tau}}{dx} \right)+G^{(2)}_1(x/z)+G^{(3)}_1(x/z)}{\frac{x^2}{z^2}J(x/z)+Y(x/z)+\frac{2\pi^2}{15}+G^{(2)}_2(x/z)+G^{(3)}_2(x/z)},
\end{align} 
where
\begin{align}
G_1^{(2)}(\omega)&= 2\pi\alpha \left[\frac{1}{\omega}\biggl(\frac{K(\omega)}{3} +2K(\omega)^2-\frac{J(\omega)}{6}-K(\omega)J(\omega)\right) \nonumber \\
&\ \ \ \ \ \ \ \ +\left(\frac{K'(\omega)}{6}-K(\omega)K'(\omega)+\frac{J'(\omega)}{6}+J'(\omega)K(\omega)+J(\omega)K'(\omega) \right) \biggl], \nonumber \\
G_2^{(2)}(\omega)&=-8\pi\alpha \left(\frac{K(\omega)}{6}+\frac{J(\omega)}{6}-\frac{1}{2}K(\omega)^2+K(\omega)J(\omega) \right) \nonumber \\
&\ \ \ \ \ \ \ \ +2\pi\alpha\omega \left(\frac{K'(\omega)}{6}-K(\omega)K'(\omega)+\frac{J'(\omega)}{6} +J'(\omega)K(\omega)+J(\omega)K'(\omega) \right)  \nonumber \\
G_1^{(3)}(\omega)&=\frac{e^3}{4\pi}\left(K(\omega)+\frac{\omega^2}{2}k(\omega) \right)^{1/2} \biggl[ \frac{1}{\omega} \left(2J(\omega)-4K(\omega) \right)-2J'(\omega)-\omega^2j'(\omega) \nonumber \\
&\ \ \ \ \ \ \ \ \ \ \ \ \ \ \ \ \ \ \ \ \ \   -\omega\left(2k(\omega)+j(\omega)\right)-\frac{\left(2J(\omega)+\omega^2j(\omega)\right)\left(\omega\left(k(\omega)-j(\omega)\right)+K'(\omega)\right)}{2\left(2K+\omega^2k(\omega)\right)} \biggl] \nonumber \\
G_2^{(3)}(\omega)&=\frac{e^3}{4\pi}\left(K(\omega)+\frac{\omega^2}{2}k(\omega) \right)^{1/2}\left[\frac{(2J(\omega)+\omega^2j(\omega))^2}{2(2K(\omega)+\omega^2k(\omega))}-\frac{2}{\omega}Y'(\omega)-\omega\left(3J'(\omega)+\omega^2j'(\omega) \right) \right] \nonumber \\
\end{align} 
with
\begin{align}
K(\omega)&=\frac{1}{\pi^2}\int^{\infty}_0du~\frac{u^2}{\sqrt{u^2+\omega^2}}\frac{1}{\exp\left(\sqrt{u^2+\omega^2}\right)+1}, \nonumber \\
J(\omega)&=\frac{1}{\pi^2}\int^{\infty}_0du~u^2\frac{\exp\left(\sqrt{u^2+\omega^2}\right)}{\left(\exp\left(\sqrt{u^2+\omega^2}\right)+1\right)^2}, \nonumber \\
Y(\omega)&=\frac{1}{\pi^2}\int^{\infty}_0du~u^4\frac{\exp\left(\sqrt{u^2+\omega^2}\right)}{\left(\exp\left(\sqrt{u^2+\omega^2}\right)+1\right)^2}, \nonumber \\
k(\omega)&=\frac{1}{\pi^2}\int^{\infty}_0du \frac{1}{\sqrt{u^2+\omega^2}}\frac{1}{\exp\left(\sqrt{u^2+\omega^2}\right)+1}, \nonumber \\
j(\omega)&=\frac{1}{\pi^2}\int^{\infty}_0du \frac{\exp\left(\sqrt{u^2+\omega^2}\right)}{\left(\exp\left(\sqrt{u^2+\omega^2}\right)+1\right)^2}.
\end{align}
The prime represents the derivative with respect to $\omega$.
Note that $G^{(2)}$ and $G^{(3)}$ indicate the finite temperature corrections at $\mathcal{O}(e^2)$ and $\mathcal{O}(e^3)$ respectively.

%%%%%%%%%%%%%%%%%%%%%%%%%%%%%%%%%%%%%%%%%%%%%%%%%%%%%%%%%%%%%%%%%%%%%%%%%%%%%%
%%%%%%%%%%%%%%%%%%%%%%%%%%%%%%%%%%%%%%%%%%%%%%%%%%%%%%%%%%%%%%%%%%%%%%%%%%%%%%

\section{Analytic estimation of the collision integral}
\label{appb}
In this appendix, we analytically perform seven out of nine integrations in the collision terms for four-Fermi interaction processes in the isotropic Universe, following refs.~\cite{Dolgov:1997mb,Blaschke:2016xxt}.  We consider the general form of the collision term in this case,
\begin{align}
C_{\rm coll}=\frac{1}{2E_1}\int(2\pi)^4\delta^{4}(\sum_ip_i)\mathcal{|M|}^2F\left(\rho_p\right)\prod_{i=2}^4\frac{d^3\bm{p}_i}{(2\pi)^32E_i},
\label{CC}
\end{align}
where $E_i$ is the energy of $i$-th particle. The matrix $F\left(\rho_p \right)$ is a function of neutrino density matrix and $|\mathcal{M}|^2$ is a part of the possible squared amplitudes summed over spin degrees of freedom of all particles except for the first particle. We use the following relation:
\begin{align}
\delta^{(3)}(\sum_i \bm{p}_i)=\int e^{\bm{\lambda}\cdot(\bm{p}_1+\bm{p}_2-\bm{p}_3-\bm{p}_4)}\frac{d^3\bm{\lambda}}{(2\pi)^3},
\label{delta}
\end{align}
and decompose momentum integrations into the radial integration and the angle integrations,
\begin{align}
d^3\bm{p}_i=p_i^2dp_i\sin\theta_id\theta_id\phi_i \equiv p_i^2dp_id\Omega_i.
\label{d^3p}
\end{align}
Using Eqs.~(\ref{delta}) and (\ref{d^3p}), we write the general collision term (\ref{CC}) as
\begin{align}
C_{\rm coll}=\frac{1}{64\pi^3E_1p_1}\int \delta(E_1+E_2-E_3-E_4)F(\rho_p(t))D(p_1,p_2,p_3,p_4)\frac{p_2dp_2}{E_2}\frac{p_3dp_3}{E_3}\frac{p_4dp_4}{E_4},
\end{align}
where
\begin{align}
D(p_1,p_2,p_3,p_4)&=\frac{p_1p_2p_3p_4}{64\pi^5}\int^{\infty}_0\lambda^2d\lambda \int e^{i\bm{\lambda}\cdot \bm{p}_1}d\Omega_\lambda\int e^{i\bm{\lambda}\cdot \bm{p}_2}d\Omega_{p_2} \nonumber \\
&\ \ \ \ \ \ \ \ \ \ \ \times \int e^{-i\bm{\lambda}\cdot \bm{p}_3}d\Omega_{p_3}\int e^{-i\bm{\lambda}\cdot \bm{p}_4}d\Omega_{p_4}|\mathcal{M}|^2.
\label{Dlambda}
\end{align}
In the cases of four-Fermi interaction processes, all of $|\mathcal{M}|^2$ have two kinds of forms,
\begin{align}
K_1({q_1}_\mu q_2^\mu)({q_3}_{\nu}q_4^\nu)&=K_1(E_1E_2-\bm{q}_1\cdot\bm{q}_2)(E_3E_4-\bm{q}_3\cdot\bm{q}_4), \label{K1} \\
K_2m^2({q_3}_\mu q_4^\mu)&=K_2m^2(E_3E_4-\bm{q}_3\cdot \bm{q}_4),
\label{K2}
\end{align}
where $q_i$ corresponds to one of $p_j$ and the angle between $\bm{q}_i$ and $\bm{q}_j$ is written in terms of the integration variables of angle,
\begin{align}
\cos\psi_{ij}=\sin\theta_i\sin\theta_j\cos(\phi_i-\phi_j)+\cos\theta_i\cos\theta_j.
\end{align}
In both cases of Eqs.~(\ref{K1}) and (\ref{K2}), we can perform all angle integrals in Eq.~(\ref{Dlambda}) so that we can write $D(q_1,q_2,q_3,q_4)$ in the case of Eq.~(\ref{K1}) as 
\begin{align}
D=K_1[E_1E_2E_3E_4D_1 + E_1E_2D_2(q_3,q_4)+E_3E_4D_2(q_1,q_2) + D_3],
\end{align}
while in the case of Eq.~(\ref{K2}),  $D(q_1,q_2,q_3,q_4)$ is expressed as
\begin{align}
D=K_2E_1E_2[E_3E_4D_1+D_2(q_3,q_4)].
\end{align}
Here $D_{1,2,3}$ are defined in Eq.~(\ref{Dfunc}) and hereafter we only consider $D_1,D_2(q_3,q_4), D_3$.

Although we can perform the integrals in $D_{1,2,3}$ and get the exact expressions given in ref.~{\cite{Blaschke:2016xxt}}, we assume for simplicity that $q_1>q_2$ and $q_3>q_4$ without loss of generality. 
Then we get the simplified expressions of $D_{1,2,3}$ in four different cases: \\
\\
$(1)\ q_1+q_2>q_3+q_4$, $q_1+q_4>q_2+q_3$ and $q_1 \leq q_2+q_3+q_4$
\begin{align}
D_1&=\frac{1}{2}(q_2+q_3+q_4-q_1), \nonumber \\
D_2(q_3,q_4)&=\frac{1}{12}\left((q_1-q_2)^3+2(q_3^3+q_4^3)-3(q_1-q_2)(q_3^2+q_4^2) \right), \nonumber \\
D_3&=\frac{1}{60}\bigl(q_1^5-5q_1^3q_2^2+5q_1^2q_2^3-q_2^5 \nonumber \\
&-5q_1^3q_3^2+5q_2^3q_3^2+5q_1^2q_3^3+5q_2^2q_3^3-q_3^5 \nonumber \\
&-5q_1^3q_4^2+5q_2^3q_4^2+5q_3^3q_4^2+5q_1^2q_4^3+5q_2^2q_4^3+5q_3^2q_4^3-q_4^5\bigl).
\label{Case1}
\end{align}
Note that the case $q_1>q_2+q_3+q_4$ is unphysical so that $D_1=D_2=D_3=0$ in this case.
\\
\\
$(2)\ q_1+q_2>q_3+q_4$ and $q_1+q_4<q_2+q_3$
\begin{align}
D_1&=q_4, \nonumber \\
D_2(q_3,q_4)&=\frac{1}{3}q_4^3, \nonumber \\
D_3&=\frac{1}{30}q_4^3\left(5q_1^2+5q_2^2+5q_3^2-q_4^2\right).
\end{align}
\\
$(3)\ q_1+q_2<q_3+q_4$, $q_1+q_4<q_2+q_3$ and $q_3 \leq q_1+q_2+q_4$
\begin{align}
D_1&=\frac{1}{2}(q_1+q_2+q_4-q_3), \nonumber \\
D_2(q_3,q_4)&=\frac{1}{12}\left(-(q_1+q_2)^3-2q_3^3+2q_4^3+3(q_1+q_2)(q_3^3+q_4^3)\right).
\end{align}
$D_3$ is equal to that in Eq.~(\ref{Case1}) with the replacement of variables $q_1 \leftrightarrow q_3$ and $q_2 \leftrightarrow q_4$ and the case $q_3>q_1+q_2+q_4$ is unphysical so that $D_1=D_2=D_3=0$ in this case.
\\
\\
$(4)\ q_1+q_2<q_3+q_4$ and $q_1+q_4>q_2+q_3$
\begin{align}
D_1&=q_2, \nonumber \\
D_2(q_3,q_4)&=\frac{1}{6}q_2\left(3q_3^2+3q_4^2-3q_1^2-q_2^2\right), \nonumber \\
D_3&=\frac{1}{30}q_2^3\left(5q_1^2+5q_3^2+5q_4^2-q_2^2 \right).
\end{align}
After we have integrated the $\delta$-function, we get the simplified expression of the collision term, leaving two integrals,
\begin{align}
C_{\rm coll}=\frac{1}{64\pi^3E_1p_1}\int \int F\left(\rho_p(t)\right)D(p_1,p_2,p_3,p_4)\frac{p_2dp_2}{E_2}\frac{p_3dp_3}{E_3},
\end{align}
where $E_4=E_1+E_2-E_3$ and $p_4=\sqrt{E_4^2-m_4^2}$.

%%%%%%%%%%%%%%%%%%%%%%%%%%%%%%%%%%%%%%%%%%%%%%%%%%%%%%%%%%%%%%%%%%%%%%%%%%%%%%
 %%%%%%%%%%%%%%%%%%%%%%%%%%%%%%%%%%%%%%%%%%%%%%%%%%%%%%%%%%%%%%%%%%%%%%%%%%%%%%
%%%%%%%%%%%%%%%%%%%%%%%%%%%%%%%%%%%%%%%%%%%%%%%%%%%%%%%%%%%%%%%%%%%%%%%%%%%%%%

%\bibliography{ref}

\end{document}